\documentclass{aa}  

\usepackage{graphicx}
\usepackage[dvipsnames]{xcolor}
\usepackage{pifont}
\usepackage{txfonts}
\usepackage{multirow}
\usepackage{placeins}
\usepackage[]{hyperref}
\hypersetup{colorlinks,citecolor=blue, linkcolor=blue, urlcolor=blue, breaklinks=true}
\begin{document}

   \title{Extending the Little Red Dot population at intermediate redshift with VIPERS}


   \author{Krzysztof~Lisiecki\inst{\ref{ncbj},\ref{inst:UVa},\ref{inst:UVa2}} \and
            Krzysztof~Hryniewicz\inst{\ref{ncbj}} \and
            Francesco~Pistis\inst{\ref{ncbj}, \ref{inaf_brera}} \and
            Micha\l{}~J.~Micha\l{}owski\inst{\ref{UAM}} \and
            Aidan~P.~Cotter\inst{\ref{ncbj}} \and            
            Maciej~Koprowski\inst{\ref{MK}} \and
            Miguel~Figueira\inst{\ref{ncbj}}\and
            Agnieszka~Pollo\inst{\ref{ncbj}}\and
            Katarzyna~Ma\l{}ek\inst{\ref{ncbj}}\and
            Olga Cucciati\inst{\ref{OC}}
          }

   \institute{
   National Centre for Nuclear Research, Pasteura 7, 093, Warsaw, Poland; \email{krzysztof.lisiecki@ncbj.gov.pl}\label{ncbj}
   \and Departamento de F\'{i}sica Te\'{o}rica, At\'{o}mica y \'{O}ptica, Universidad de Valladolid, 47011 Valladolid, Spain\label{inst:UVa}
   \and Laboratory for Disruptive Interdisciplinary Science (LaDIS), Universidad de Valladolid, 47011 Valladolid, Spain\label{inst:UVa2}
   \and INAF - Osservatorio Astronomico di Brera, Via Brera 28, 20122 Milano, via E. Bianchi 46, 23807 Merate, Italy \label{inaf_brera} 
   \and Astronomical Observatory Institute, Faculty of Physics, Adam Mickiewicz University, ul. S\l{}oneczna 36, 60-286 Poznan, Poland\label{UAM}
   \and Institute of Astronomy, Faculty of Physics, Astronomy and Informatics, Nicolaus Copernicus University, Grudziadzka 5, 87-100
   Torun, Poland\label{MK}
   \and INAF - Osservatorio di Astrofisica e Scienza dello Spazio di Bologna via Gobetti 93/3 40129 Bologna, ITALY\label{OC}
   }
   \date{Received XX.XX.XXXX; accepted XX.XX.XXXX}
 
  \abstract
   {Little red dots (LRDs) are a recently identified population of compact, red sources characterised by a distinctive V-shaped UV-to-optical continuum and broad emission lines. Given their recent discovery, their physical structure and evolution remain uncertain, and their properties at intermediate redshifts are largely unexplored.}
   {We aim to identify and characterise LRD-like sources at $0.5\leq z\leq1.75$ using VIMOS Public Extragalactic Redshift Survey (VIPERS), extending the observational study to a previously poorly explored cosmic epoch. We investigate their photometric and spectroscopic properties, morphologies, abundance evolution, and local environments.}
   {We select LRD-like sources based on the UV and optical slopes estimated from broadband photometry and further refine our sample by identifying broad emission-lines from VIPERS spectroscopy. We use HSC imaging to investigate their morphology and constrain their physical sizes. We estimate their comoving number density and its evolution with redshift, and characterise their local environments using the galaxy density field of the VIPERS survey. Finally, we model the multi-wavelength emission of the unique X-ray-detected LRD-like source with CIGALE, including the X-ray constraints and active galaxy nucleus (AGN) emission.}
   {We identify 14 LRD-like sources, including one object with detected in X-rays, for which the HSC morphologies remain unresolved. We measure number density in two redshift bins ($0.5\leq z\leq1$ and $1\leq z \leq 1.75)$. We combine our results with values from literature and we find a rapid decrease in number density after cosmic noon. Two sources whose environments can be characterized are found in underdense regions. The SED of the X-ray-detected source is best reproduced by the radiatively efficient accretion disk observed at low inclination.}
   {Our results demonstrate that LRD-like sources are present at $z\geq 0.5$, filling the gap between local and high-$z$ studies. Their abundance, compact morphology, environmental properties, and multi-wavelength emission provide new constraints on the evolution and physical origin of this population.}
   \keywords{Galaxies: active – Galaxies: evolution – quasars: emission lines}
   \maketitle
%
\nolinenumbers
\section{Introduction}
The James Webb Space Telescope (JWST) has recently discovered a population of compact, high redshift sources known as little red dots \citep[LRDs; e.g.][]{Matthee2024, deGraaff2025, Wang2025}. 
They present unique spectrophotometric properties, such as V-shaped rest-frame ultraviolet-optical spectral energy distributions \citep[SEDs;][]{Kokorev2024, Kocevski2025, Akins25}, showing a strong turnover near the Balmer limit \citep{Setton25}.
In addition to the SED shape, LRDs exhibit broad Balmer emission lines, lack of variability, and X-ray quietness \citep[e.g.][]{Kokorev2024,Hviding2025, Kocevski2025, Kokubo2025, Liu2026}.
LRDs are also extremely compact in rest-frame optical images, unresolved with JWST, which indicates sizes of $\leq100-300$pc \citep[e.g.][]{Baggen24, Tripodi25, Cloonan26, Whalen26}, with the exception of one LRD-like object found within a resolved spiral galaxy \citep{Rinaldi26}.
The combination of these properties makes LRDs different from conventional unobscured quasars and ordinary star-forming galaxies.

So far, standard models of active galactic nuclei (AGN) and stellar populations have failed to explain the observed SEDs of LRDs \citep[e.g.][]{Killi24, Perez-gonzalez24}.
Some studies argue that the properties can be explained with super-Eddington accretion \citep[e.g.][]{Madau26}, or mixed AGN with extreme starburst \citep[][]{Perez-gonzalez24}.
However, recent studies show that the LRDs could represent a short phase of supermassive black hole (SMBH) growth.
In such scenario, the central engine is embedded in cocoon-like structure of dense, compact, and highly-ionised gas reservoir, explaining the observed features \citep[][]{deGraaff2025, Nandal26, Rusakov26, Sneppen26a, Sneppen26b}.
Consequently, determining whether LRDs represent a short-lived phase of black-hole growth, a particular geometry of accretion, a composite AGN–starburst population, or several physically distinct populations, requires more detailed observations, which are not easily obtained from high-redshift samples.

Since LRDs were discovered with JWST, the focus has mainly been on the early Universe, with particularly large samples at $z>4$.
Recent studies have reported samples of LRDs at intermediate \citep[][]{Ma26, Weibel26} and low redshift \citep{Ding26,Ji2026, Lin2026a,Lin2026b,Park2026}, suggesting a strong evolution of their number densities. 
The rapid decline toward lower-$z$ is interpreted as evidence that LRDs represent a short-lived evolutionary phase, which should result in another class of objects, representing the late stage of an LRD.
Recently \cite{Fu26} reported two objects, referred to as Forges, that exhibit the transitional properties between LRD and quasi stellar object (QSO).
In addition, \cite{Hviding26} studied an X-ray-detected LRD (XRD), interpreting it as the late stage cocoon around the SMBH.

If LRDs represent rapidly growing SMBHs embedded in dense gas cocoons, their formation may require an efficient supply of gas to the nuclear region.
Such gas inflows may be facilitated by galaxy–galaxy interactions and mergers, and potentially linked to the larger-scale environment. 
The local overdensities in the cosmic web could provide favourable conditions for the formation of the compact, gas-rich nuclear structures required to sustain rapid black-hole growth.
Conversely, if the LRD phase represents a short-lived stage of individual galaxy evolution, its occurrence may not be strongly associated with an overdense environment.
Observational constraints on the environments of LRDs remain inconclusive: \cite{Carranza-Escudero25} find that LRDs do not preferentially inhabit overdense regions, whereas \cite{Schindler25} identify a spectroscopically confirmed LRD embedded in a prominent galaxy overdensity.
\cite{Zhang26} claimed that the LRD-like objects at $z<3.5$ tend to occupy environments comparable of those of ordinary galaxies.
Another recent case study further suggests that galaxy interactions may trigger or regulate an LRD-like phase through enhanced gas accretion \citep[][]{Merida26}. 
Therefore, measurements of the local environments of LRDs provide an independent test of their formation and evolutionary scenarios. Extending such measurements to lower redshift using a complete and homogeneous galaxy density field can provide a valuable constraint on whether the environmental conditions associated with the LRD phenomenon evolve over cosmic time.

In this work we aim to identify LRD-like objects within the VIMOS Public Extragalactic Redshift Survey (VIPERS), combining their broad-band continuum slopes and spectroscopic properties. 
We measure the comoving number density of LRDs and investigate its evolution over cosmic time.
Additionally, we characterise their local environments with a local density field. 
Finally, we perform multi-wavelength SED modeling of an X-ray-detected LRD-like source, to investigate whether an AGN embedded in an obscured/reprocessing medium can reproduce its observed properties.
Together, these analyses provide a multi-dimensional view of LRD-like objects at $z>0.5$.

Throughout the paper, we adopt the \cite{Chabrier03} stellar initial mass function and we assume the $\Lambda$CDM cosmological model with $H_0$ = 70 km/s/Mpc, $\Omega_m$ = 0.3, and $\Omega_\Lambda$ = 0.7.

\section{Data} \label{sec:data}
\subsection{VIPERS spectroscopy}
The VIMOS Public Extragalactic Redshift Survey (VIPERS) is a completed ESO Large Program designed to investigate the spatial distribution of galaxies over the redshift range $0.5\leq z\leq1.2$ \citep{garilli14,guzzo14, scodeggio18}. 
VIPERS provides a spectroscopic catalogue for more than 90\,000 sources selected within the W1 and W4 fields of the Canada-France-Hawaii Telescope Legacy Survey (CFHTLS) Wide.
Targets were selected to a limit of $i_\textrm{AB} < 22.5$~mag with simple and robust colour-colour pre-selection ($r-i$ vs $u-g$) to effectively remove stellar contaminants and galaxies outside the redshift range of interest \citep[see][for details]{guzzo14}.
Extending over an area of $23.5$ deg$^2$, VIPERS can be considered as the intermediate redshift ($z\sim0.7$) equivalent of state-of-the-art local surveys ($z<0.5$), such as Sloan Digital Sky Survey \citep[SDSS,][]{york00, ahmuda20, SDSS26}, or Galaxy and Mass Assembly \citep[GAMA,][]{Driver11}.
The spectra were observed using the VIMOS spectrograph \citep{lefevre03} with the LR Red grism and resulted with a wavelength coverage of 5\,500-9\,500 $\AA$ with a resolution R$ \simeq 220$.
A detailed description of the survey is given by \citet{guzzo14} and \citet{scodeggio18}, and the specifications of the pipeline used for data reduction with the quality flag system are described by \citet{garilli14}.

In the following analysis, in particular the number density estimations, we used the spectroscopic success rate (SSR), and the target sampling rate (TSR). 
The SSR is a ratio between the number of objects for which VIPERS has successfully measured a redshift and the number of objects targeted by the spectroscopic observations, effectively tracing the probability of a target source to get a reliable spectrum.
The TSR is a ratio of the local surface densities of target and parent galaxies, measuring the probability of a source to be selected as a target.
Both parameters measure the completeness of galaxies within the VIPERS catalogue (see \citealt{garilli14} and Section 1 of \citealt{scodeggio18}).

\subsection{Main photometric catalogue}
The VIPERS spectroscopic results, such as spectroscopic redshifts, are complemented by ancillary information.
We especially leverage the multiwavelength photometric catalogue, taken from the VIPERS database \citep{moutard16_I}, which contains observations from ultraviolet (UV) to mid-infrared (MIR) wavelengths. 
Throughout this study, we use the following bands from the catalogue: far-UV (FUV) and near-UV (NUV) from Galaxy Evolution Explorer (GALEX); \textit{u, g, r, i, z, y} from CFHTLS photometric measurements;  K$_s$ from the CFHT Wide-field Infrared Camera (WIRCam); K from Visible and Infrared Survey Telescope for Astronomy (VISTA) Deep Extragalactic Observations (VIDEO) survey \citep{jarvis13};  W1, W2, W3, and W4 passbands (3.4, 4.6, 12.1, and 22.5~$\mu$m) from NASA’s Wide-field Infrared Survey Explorer \cite[WISE;][]{wright10}.

\subsection{Auxiliary data}
\subsubsection{Hyper Suprime-Cam Subaru Strategic Program}
Hyper Suprime-Cam Subaru Strategic Program (HSC-SSP, \citealt{Aihara22}) is a multiwavelength observational program covering over 1400 deg$^2$.
The imaging in five broad-band filters, \textit{g, r, i, z} and \textit{y}, was done in three modes: Wide, Deep and Ultradeep.
Depending on the mode and the band, the depth of the imaging varies from 24.4 to 28.1 mag\footnote{\url{hsc.mtk.nao.ac.jp/ssp/}}.
As the ultradeep/deep mode is not available for all of our galaxies, we use the wide mode instead, ensuring the consistency and homogeneity of the selection. 

We use HSC-SSP data for the morphology analysis only as it is the best available option for such measurements due to its depth and spatial resolution. 
For each source of interest, we cut square stamps of size 20$"$ centred on the sources.
The PSFs for each observation are available via online, dedicated tools\footnote{\url{hsc-release.mtk.nao.ac.jp/doc/}}.

\subsubsection{GALEX}
The rest-frame UV coverage is crucial for $\beta_{\mbox{uv}}$ slope estimation ($F_\lambda \propto \lambda^{\beta_{\mbox{uv}}}$).
Thus, we crossmatch galaxies that do not have GALEX data points in the catalogue thereby not allowing to constrain $\beta_{\mbox{uv}}$ slope with GALEX Data Release 6 and 7 of the All-Sky Imaging Survey\footnote{GR6+7 AIS} catalogue \citep[][]{Bianchi14}. 
We searched for the corresponding sources within a distance of 2 arcseconds, resulting in additional measurements for 74 galaxies.

Additionally, we performed photometry in GALEX observations for that were also not present in GR6+7 AIS catalogue.
The detailed description of the analysis can be found in the Appendix \ref{app:galex}.
In short, we download the GALEX patches with sources of interest from MAST and measured the fluxes with both S\'ersic modelling and aperture (4.5 or 7.5 arcsec) and report the mean value. 
This resulted in additional measurements for 11 VIPERS broad line galaxies.

\subsubsection{XMM-Newton}
As LRDs are observed to be X-ray suppressed, and it may be a separating factor between LRDs, XRDs or Forge-like objects, we cross-match our final sample (see Section~\ref{sec:unresolvedmorphology}) with 5XMM-DR15 catalogue \citep[][]{Webb26}.
Considering only counterparts with separation $<2"$, we find one X-ray detection for our main LRD sample.

\subsubsection{HELP}
To better constrain the SED of the X-ray detected source, we cross match it with the Herschel Extragalactic Legacy Project \citep[HELP;][]{Shirley21}.
We use fluxes reported in the HELP catalogue, specifically: 2MASS J and H bands \citep[][]{Huchra12};  3.6, 4.5, 5.8, and 8~$\mu$m channels from \textit{Spitzer}/Infrared Array Camera; MIPS 24$\mu$m band; Herschel PACS \citep[][]{Poglitsch10} green ($\sim100\mu$m and red ($\sim160\mu$m) bands; and Herschel SPIRE \citep[][]{Griffin10} 250, 350, 500 $\mu$m bands.

\subsubsection{Local densities}\label{sec:datadensity}
The local environment of each galaxy at $z\leq1$ is characterised using the VIPERS density contrast ($\delta$), as:
\begin{equation}
    \delta(\textrm{Ra, Dec, } z) = \frac{\rho(\textrm{Ra, Dec, } z) - \overline{\rho(z)}}{\overline{\rho(z)}},
\end{equation}
where $\rho(\textrm{Ra, Dec, } z)$ is the local galaxy density at the position of each source and $\overline{\rho(z)}$ is the mean galaxy density at the corresponding redshift.
The local density is estimated by applying a cylindrical top-hat filter whose projected radius is set by the distance to the fifth nearest neighbour, while the cylinder extends over $\pm1000~\mathrm{km,s^{-1}}$ along the line of sight to reduce the impact of peculiar velocities. 
The density field is traced by a volume-limited galaxy sample selected according to $M_B \leq -20.4-z$, which provides a nearly constant comoving number density and is complete up to ($z=0.9$). 
A detailed description of the derivation of local densities can be found in \citet[][]{Cucciati14, Cucciati17}.

\section{Sample selection}\label{sec:sample}
In this section, we describe all steps to select LRDs within the VIPERS.
The steps are summarised in Table~\ref{Tab:selection}.

\begin{table}[]
    \centering
    \caption{Summary of our selection.}
    \begin{tabular}{lcc}
         \hline
         Cut & Number of sources & Percentage  \\
         \hline
         VIPERS & 92\,520 & 100.000\%\\
         Broad line flag & \phantom{0}1\,247 & \phantom{00}1.348\%\\
         Confident $z_\textrm{spec}$ flag & \phantom{0}1\,163 & \phantom{00}1.257\%\\
         $z>0.5$ & \phantom{0}1\,128 & \phantom{00}1.219\%\\
         V-shape & \phantom{00\,0}75 & \phantom{00}0.081\%\\
         Unresolved & \phantom{00\,0}15 & \phantom{00}0.016\%\\
         Undetected [\ion{Ne}{V}] & \phantom{00\,0}14&\phantom{00}0.015\%\\
         \hline
         V-LRDs & \phantom{00\,0}13 &\phantom{00}0.014\%\\
         X-LRD & \phantom{00\,00}1 & \phantom{00}0.001\%\\
         \hline
    \end{tabular}
    \label{Tab:selection}
\end{table}

\subsection{Preselection}
We start with sources with visible broad line emission.
For this purpose, we utilise the VIPERS quality flags.
The quality flags were used to indicate the redshift estimation quality and the broad line component with human visual validation \citep[][]{scodeggio18}.
We further discard sources without a secure redshift estimate, with confidence level <90\%.
Finally, we focus only on sources at $z\geq0.5$, as VIPERS was designed to observed galaxies above redshift 0.5, and any other detections can be produced by peculiar objects.  

\subsection{V-shape}

To select only the characteristic V-shape SED, we look for UV upturn via rest-frame $\beta_{\mbox{uv}}$ (UV slope).
Since for most of the sources, no spectra in rest frame UV is available, we utilize broad band photometry.
According to \cite{Setton25}, the break in the SED should occur close to 3\,645$\AA$, which is the Balmer limit.
Thus, we defined $\beta_{\mbox{uv}}$ as the slope of broad band observations between 1\,200-3\,200 {\AA} in rest frame in log-log space: $\log(F_\lambda) \propto \beta_{\mbox{uv}}\log(\lambda)$.

\begin{figure}[]
\includegraphics[width = 0.48\textwidth]{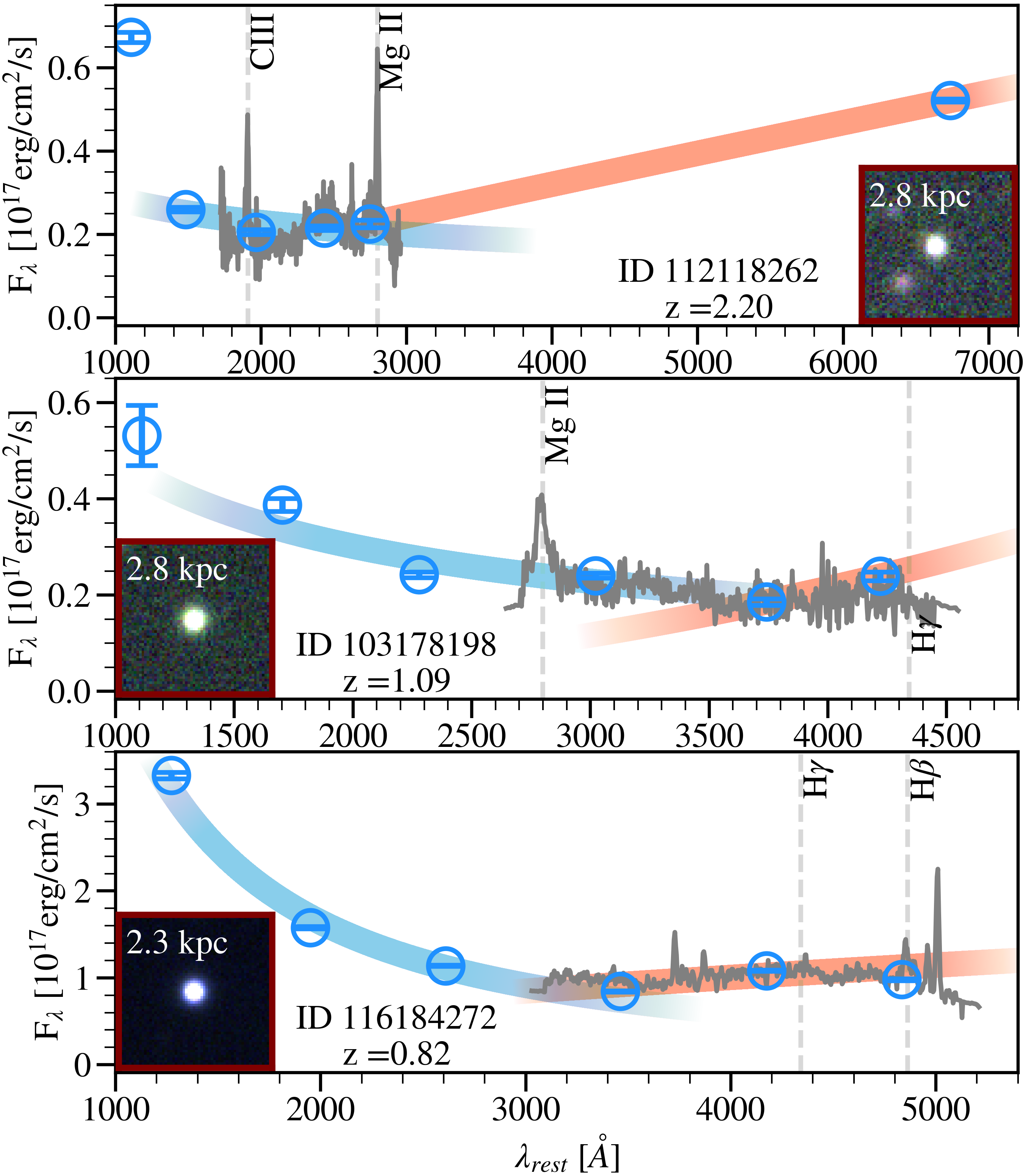}
\caption{Examples of rest frame SEDs and cutouts of our candidates through redshift. The solid gray line shows the scaled spectrum, the blue points represent the broadband observations with errors. 
We plot the optical and UV slopes with light-red and light-blue, respectively. 
We mark the lines considered to be broad with light-gray dashed line. Finally, we provide the 10$\times$10 arcsec cutouts. The half-light radius (resolution limit) in kpc of the PSF in $i$-band is denoted in the cutout. The SEDs of all objects in the final sample are presented in Figure~\ref{fig:sed1}.}
\label{fig:exampleSED}
\end{figure}

We also test for positive optical slope. 
However, since data are inhomogeneous (i.e. sometimes there is only one broad band observation between 3\,200-8\,000 \AA), fitting is not always possible. 
Where we can, we fit the slope to all observations between 3\,200-8\,000~$\AA$.
Otherwise, we measure the optical slope using the reddest point (observation with the largest $\lambda$) used to estimate the $\beta_{\mbox{\mbox{uv}}}$ and single available data-point in the range 3\,200-8\,000 \AA (see upper panel of Figure~\ref{fig:exampleSED}).

Finally, to select V-shaped objects, we follow the same criteria as in previous studies \citep[e.g.][]{Kocevski2025, Ding26, Seyberlich26}, i.e. we require $\beta_{\mbox{\mbox{uv}}}<-0.25$ and positive optical slope.
We find 75 V-shaped sources.
We present a few examples of V-shaped SEDs in Figure~\ref{fig:exampleSED}.

\subsection{Unresolved morphology}\label{sec:unresolvedmorphology}

One source of potential contaminates with simple V-shape selection would be the post starburst galaxies \citep[][]{Perez-gonzalez24}.
However, the classical LRDs are point-like objects with JWST resolution, sometimes reaching upper limits of few tens of pc. 
Our observations have much worse spatial resolution compared to JWST.
Any object of such little size will be unresolved at $z\geq0.5$.
Thus, to separate potential LRDs from post starburst galaxies, we further look for only unresolved objects.
For this purpose, we study the morphologies of V-shaped objects.
The detailed description of our fitting and error estimation can be found in the Appendix~\ref{app:morpho}.
We perform two GALFIT runs: 1)~simple point source (PSF-like model); 2) two-component model consisting of the S\'ersic profile combined with point source.
We performed GALFIT fits for each galaxy with V-shape SED in two HSC bands: \textit{i}-band and \textit{y}-band.
We use \textit{i}-band since it has the best resolution (median seeing$\sim0.6''$) and depth (26.2~mag), while the y-band is probing the optical light towards higher-$z$ sources.
Our analysis results not only in the best fit solution, but also in the robust uncertainties. 
This approach allows us to decouple the host from point-like AGN.

\begin{figure}[]
\includegraphics[width = 0.48\textwidth]{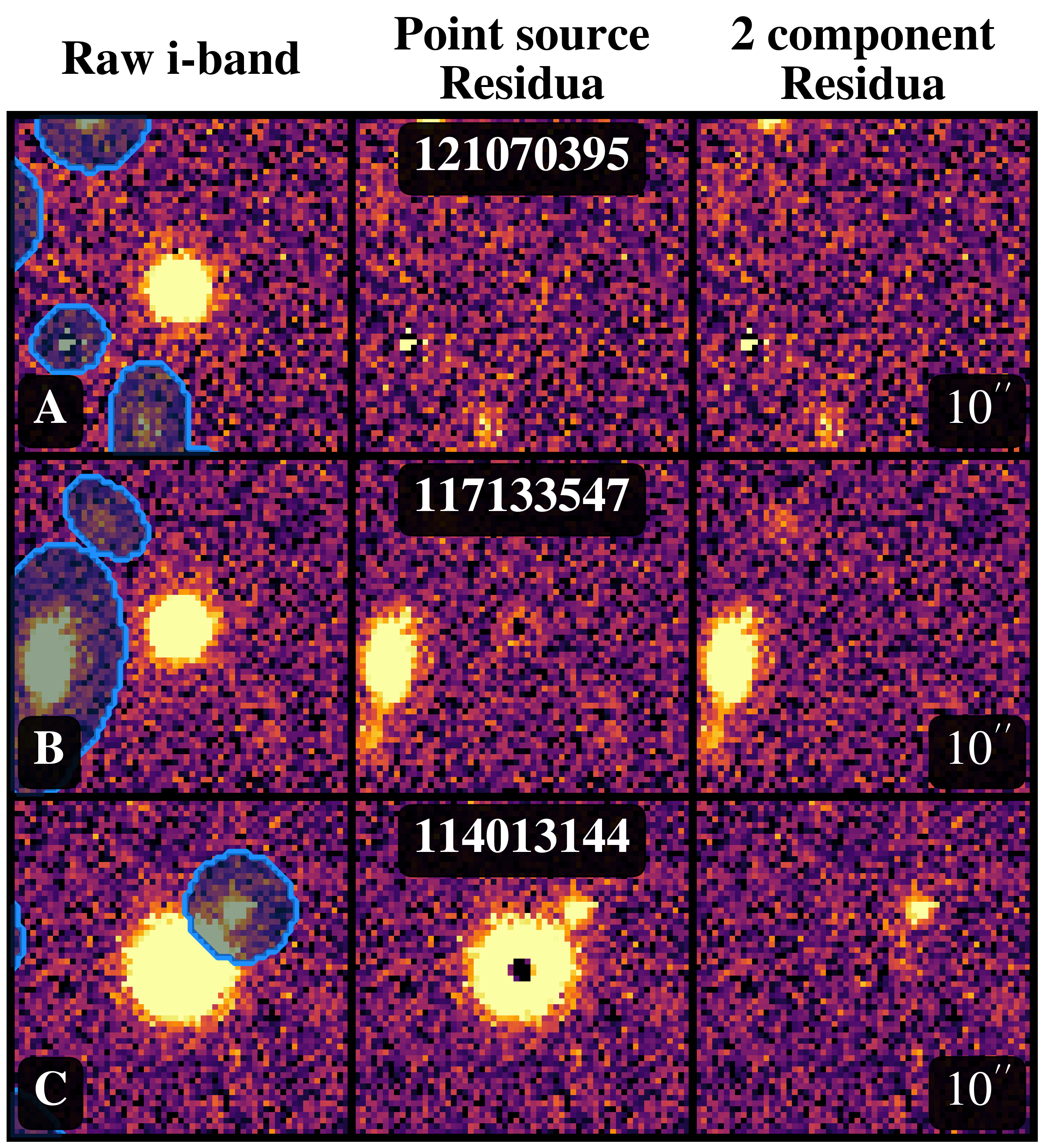}
\caption{Examples of GALFIT fits following our methodology. We show the raw i-band in the left column, the residua from point source model in the middle column and residua from 2 component model in the right one. We present the mask in blue. The VIPERS ID is given for each source in the middle column. The rows from top: \textit{A}) point source model sufficient; \textit{B}) point source not enough, but second criterion holds; \textit{C}) resolved and removed from the sample.}
\label{fig:galfitFits}
\end{figure}

We then use a 2-step criterion to decide if the source is considered unresolved. (1)~We compare the Bayesian Information Criteria (BIC) of point-like fit  and two-component fit. If the BIC of point-like model is lower in both bands we consider a source unresolved. There are 3 galaxies that meet this criterion.
(2) Otherwise, we find the circularised effective radius of the S\'ersic profile in two-component fit, defined as $R_e = a\sqrt{q}$, where $a$ is major axis and $q$ is the ratio of minor to major axis, with corresponding uncertainty $\Delta R_e$.
We measure the half-light radius of the PSF profile\footnote{This is the conversion between FWHM and half-light radius for a Gaussian profile. This is a more conservative approach than measuring the half-light radius of PSF directly.} $R_{e}^ \textrm{psf} = \mbox{FWHM}^\textrm{psf}/(2)$, where FWHM$^\textrm{psf}$ is the full width half maximum of the PSF profile. 
Finally, we test whether $R_e+\Delta R_e<R_{e}^ \textrm{psf}$ for both bands (the result is at least $1\sigma$ from resolved scenario). 
There are additional 12 galaxies that meet this criterion.  
Examples of different outcomes of our criteria are presented in Figure~\ref{fig:galfitFits}.
This leaves us with a sample consisting of 15 sources with broad lines, V-shape SED, and unresolved morphology.

It should be noted that some studies suggest that LRDs are more extended in UV than in optical \citep[e.g.][]{Cloonan26}.
The \textit{i} and \textit{y}-bands are probing optical light up to $z\sim1.15$.
Above this threshold, \textit{i}-band increasingly probes the rest-frame UV.
Requiring compact morphology in both bands may therefore preferentially exclude LRDs with extended UV emission but unresolved optical counterparts. 
Thus, we checked if any objects would enter our selection, if for z>1.15 we would require only HSC-y to be unresolved. 
We report that the final sample would not change.

\subsection{Final sample}
As a final selection criterion, we aim to reduce potential contamination by conventional broad-line AGNs exhibiting strong high-ionisation emission lines. 
LRDs spectra are often characterised by weak or absent high-ionisation transitions, including [\ion{Ne}{V}]\,$\lambda3427\AA$ \citep[e.g.][]{Park2026}. 
The presence of [\ion{Ne}{V}], which requires photons with energies exceeding approximately 97 eV, is therefore used as an additional empirical discriminator.
We perform spectral fitting of our sources (see Appendix~\ref{sec:spectralfitting}) and exclude one object with a measured [\ion{Ne}{V}] emission line satisfying $\textrm{SNR}_\textrm{ [\ion{Ne}{V}]}>2$ and $\textrm{EW}_\textrm{[\ion{Ne}{V}]}>3$\AA. 
We emphasise that this criterion does not provide an unambiguous separation between LRDs and conventional AGNs, as [\ion{Ne}{V}] is not detected in all AGNs \citep{Reiss25}.
It serves to refine the purity of our sample.

We further tested for detection in the  stacked spectrum of the remaining 14 sources.
We describe the stacking and show the spectrum in Appendix~\ref{app:stack}.
We report no detection in the stacked spectrum.
In particular, none of the individual channels nor the cumulative flux of the spectrum around the line ($\pm10~\AA$) exceeds 2$\sigma$.
Unfortunately, VIPERS spectra can only trace [\ion{Ne}{V}] $\lambda$3427 \AA \  at $0.6 \gtrsim z \gtrsim 1.8$.
While we do not have objects with $z<0.6$ in our final sample, we have one galaxy at $z>1.8$, which we keep in the sample.

Finally, we established a final sample consisting of 14 sources.
We report the observed properties of each galaxy, as well as the SEDs and cutouts, in Appendix~\ref{App:finalPresent}.
Among them, we report the detection of 1 source with X-ray counterpart observed with XMM-Newton.
From now we will refer to 13 sources without X-ray detections as V-LRDs (VIPERS little red dots), while to the remaining source as V-XRD (VIPERS X-ray dot).

\section{Results and discussion}
In this section, we describe the analysis we conducted with our final sample and compare the results with the literature.

\subsection{The SED shapes among LRDs}
\begin{figure}[]
\includegraphics[width = 0.48\textwidth]{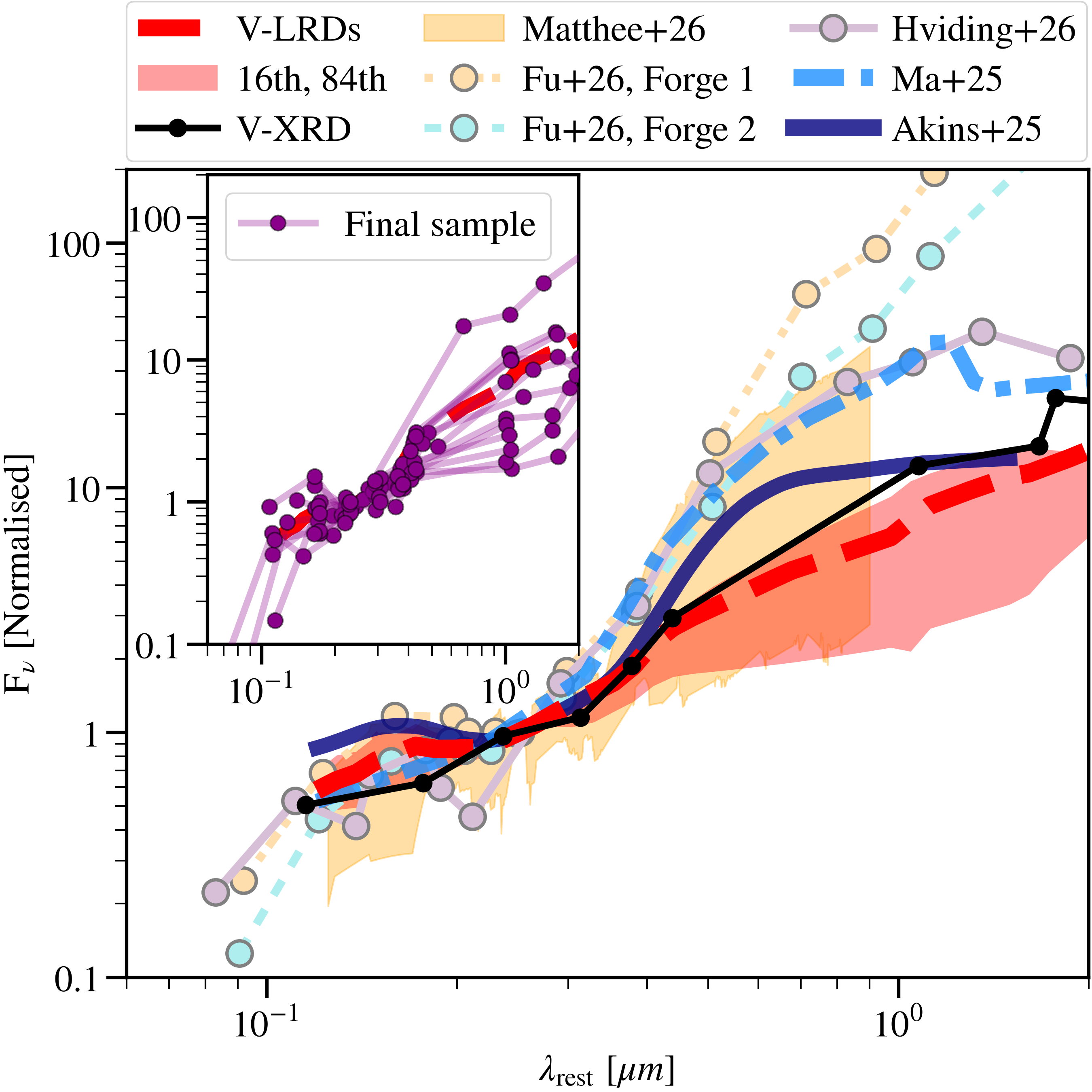}
\caption{Stacked rest frame SED of V-LRDs combined and normalised to 0.25~$\mu$m emission. The dashed red line represents the averaged trend among V-LRDs, while the red shaded area shows the 16th and 84th percentile. 
The solid black line represents the V-XRD source.
We compare our results with LRD-like objects studied by: \cite{Akins25,Fu26,Hviding26, Ma26, Matthee26}.
The inset axis show the same space with SEDs of each object in the final sample marked as purple lines. 
}
\label{fig:averageSED}
\end{figure}

As the V-shape of the SEDs is one of the most characteristic features of LRDs, we compare the shape of SEDs in our final sample with other LRD-like objects in the literature.
For this purpose, we construct a stacked SED.
We convert all photometric observations to the rest frame.
We further interpolate the SED into the common wavelength grid between 0.12-2$\mu$m and normalise them to $0.25\mu$m emission.
Finally, we calculate the averaged SED, and 16th and 84th percentiles. 
We compare the stacked SED of our final sample to: (1) the LRDs (or LRDs candidates) samples studied by \cite{Akins25} and \cite{Ma26}: (2) the three objects with X-ray detections studied by \citet[][called Forge 1, 2]{Fu26} and \citet[][called XRD]{Hviding26}; (3) and the sample of JWST/NIRSpec observed LRDs studied by \cite{Matthee26}.
Instead of showing each spectra of \cite{Matthee26} sample, we show the range of flux, similarly to \cite{Park2026}, but with masked emission lines.
The results are shown in Figure~\ref{fig:averageSED}.

The stacked SED of the V-LRDs is in agreement with the high$-z$ LRDs studied by \cite{Matthee26}, which overlap throughout the wavelength range.
The SED of the high$-z$, photometrically selected LRDs \citep[][]{Akins25} and one of the Forges \citep[][]{Fu26} exhibit slightly stronger UV excess/upturn. 
However, our results are consistent with previous studies.

The differences become more pronounced in the infrared. 
We find good agreement with the LRD spectra presented by other studies, with the mean optical slope of the sources in our final sample, measured over $0.32<\lambda_{\rm rest}<0.8,\mu$m, being $\beta_{\rm opt}\simeq0.93$ (see Table \ref{Tab:finaltable}). 
However, our stacked SED also shows an apparent additional break at $\sim0.5,\mu$m.
This feature may be at least partly driven by the observational limitations of the available data, in particular the lack of observations over $0.5<\lambda_{\rm rest}<1,\mu$m. 
If the optical-NIR continuum remains approximately flat \citep[as the one of][]{Akins25} over this wavelength range, the lack of direct measurements may introduce artificially decreased interpolation of the SED toward shorter wavelengths, thereby producing an apparent break around $\sim0.5,\mu$m.
Nevertheless, we test for possible contamination of our sample. 
One potential contaminant is unobscured QSOs, which would generally be expected to produce detectable X-ray emission. 
This appears unlikely, however, as 9 out of 13 V-LRDs lie within the XMM-XXL North footprint and remain undetected in X-rays. 
Furthermore, excluding the four sources outside the X-ray footprint does not significantly alter the shape of the stacked SED, nor does it change its 16th percentile. 
A second potential source of contamination is normal star-forming galaxies. 
At $z\sim2.5$, the majority of galaxies are expected to have effective radii of $R_e\sim2$ kpc \citep[e.g.][]{vanderwel14, Ormerod24}, and would therefore remain unresolved in the HSC data given our methodology. 
This concern is reduced at $z<1.5$, where the mean effective radius is $\gtrsim3$ kpc. 
Nevertheless, even an unresolved star-forming galaxy would struggle to reproduce the combination of broad emission lines and strong UV emission observed in our sources, particularly in the absence of detectable X-ray emission. 
While these tests suggest that contamination is unlikely to dominate our sample, follow-up observations at higher spatial resolution will ultimately be required to robustly exclude unresolved contaminants.

\subsection{Decay of LRDs number density}
We divide our V-LRDs (13 objects) into three redshift bins: $0.5-1, 1-1.75, 1.75-2.6$.
The first bin targets the main interest of VIPERS.
The second bin range allows us to discard sources with [\ion{Ne}{V}] $\lambda$3427 \AA \  detection, as well as to probe optical light with HSC-y band. 
The last bin suffers the most from incompleteness and other biases.
We decide to cut the sample at $z=2.6$ as this is the 90th percentile of the redshift distribution for all broad-line objects in the VIPERS catalogue. 
It is important to note that at $0.5\leq z\leq1.75$ the broad line component may be observed for H$\beta$, H$\gamma$, or \ion{Mg}{II} $\lambda$2800 \AA.
Above that redshift, H$\beta$ and H$\gamma$ are out of the observed range, and \ion{Mg}{II} falls into the fringing region, making the identification of the broad-line emission difficult.
Although the spectra at this redshift may contain high-ionisation lines such as \ion{C}{IV} $\lambda$1550 \AA \  and \ion{C}{III}] $\lambda$1909 \AA, they are not prominent in LRD spectra \citep[][]{Greene24, Wang26, Fu26}.
Additionally, at $z>1.75$ there are not enough data to discard typical Type I AGNs from this bin.
Thus, we report the number density in this bin only as a lower limit and we do not interpret this bin in further analysis.

\begin{figure}[]
\includegraphics[width = 0.48\textwidth]{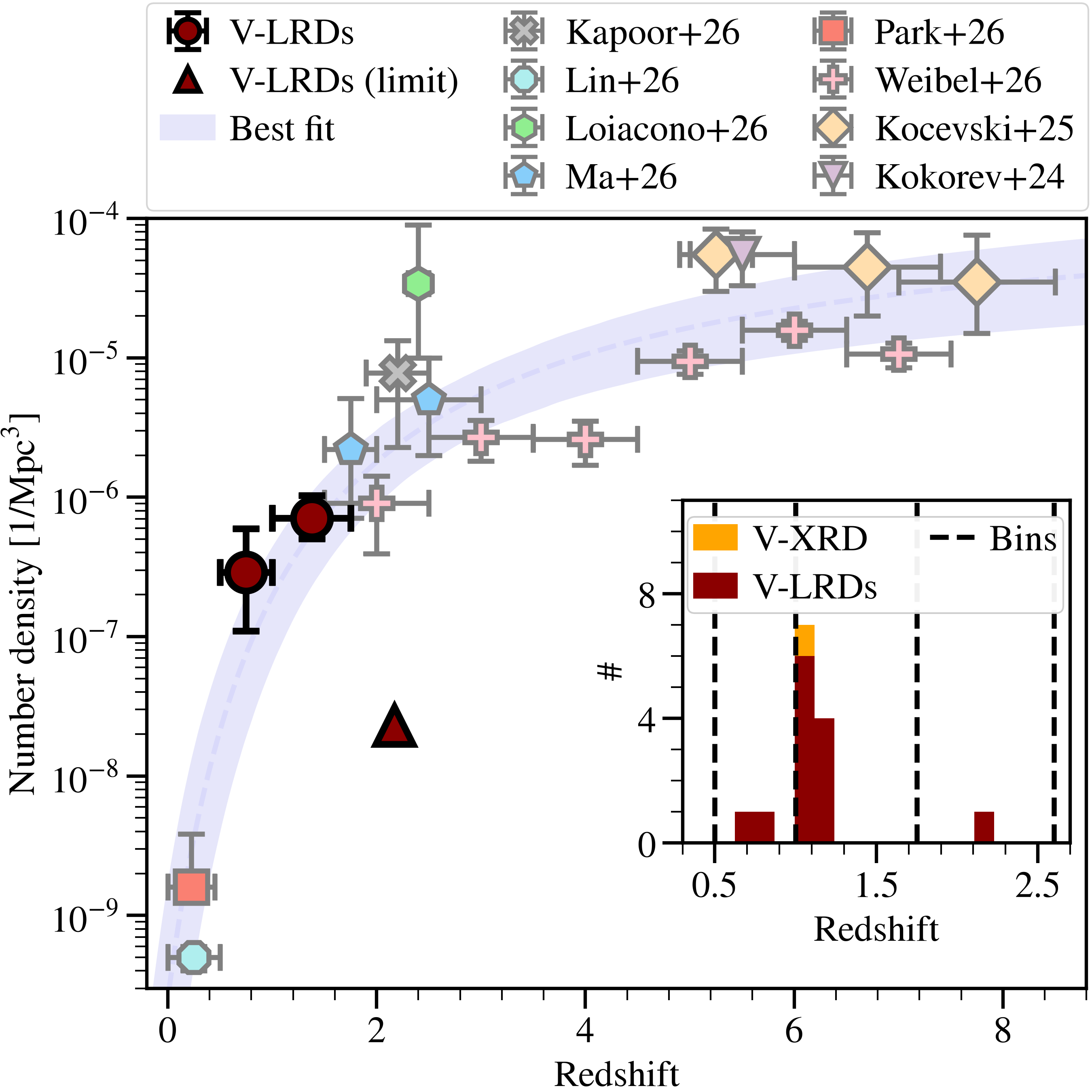}
\caption{Evolution of number density of LRDs per comoving Mpc$^3$ with redshift. We present our results (V-LRDs) with dark red points (triangle is a lower limit). We compare VIPERS LRDs to the samples studied by \cite{Kokorev2024, Kocevski2025, Kapoor26, Lin2026b, Loiacono26, Ma26, Park2026, Weibel26}.
The best fit of number density evolution with 3$\sigma$ uncertainties is presented as shaded region.
The inset axis present the redshift distribution and bin ranges of our final sample.}
\label{fig:numberDensity}
\end{figure}

For each redshift bin, we calculate the completeness-corrected number of sources, following the procedure described in Appendix~\ref{app:completeness}. 
Where available, we account for the VIPERS target sampling and spectroscopic success rates using the TSR and SSR, respectively.
For sources without these corrections, we estimate the effective observation probability empirically by comparing the number of sources in the parent and final VIPERS catalogues. 
We additionally account for the potential incompleteness caused by the exclusion of stellar contaminants, which may contain missed AGN.
Finally, we normalise the completeness-corrected counts by the surveyed area and the corresponding comoving volume to obtain the number density in units of comoving Mpc$^{-3}$. The resulting number densities are presented in Figure~\ref{fig:numberDensity} and Table~\ref{Tab:numberdensities}.

Our results fill the gap between the local Universe studied with DESI by \cite{Lin2026a} and \cite{Park2026}, and the intermediate-$z$ studied with HSC by \cite{Ma26}.
We combine all presented measurements and fit a linear function in $a - \log(N_d)$ space, where $a\equiv1/(1+z)$ is the scale factor and $N_d$ is number density in 1/comoving Mpc$^3$.
We find 
\begin{equation}
    \log(N_d) = (-5.8\pm0.5)a -3.8\pm0.1.
\end{equation}
We further reproject the relation from $a$ to redshift and present it with the $3\sigma$ range as shaded region in Figure~\ref{fig:numberDensity}.
Each measurement presented in the plot was collected in a different survey with different completeness treatment. This implied that the parameters of the derived fit should be treated as approximate, as the errors are likely underestimated.
To understand the evolution of number density of LRDs and disentangle the underlying driver, further analysis based on more physical modeling will be required.

The combined results suggest an evolution of number density, with a maximum scatter of $\sim$1 order of magnitude.
Although \cite{Loiacono26} state that there is no visible evolution of the number density at ${2.4<z<4.5}$, their result is actually in line with our fit (the distance is smaller than 2$\sigma$).
Additionally, \cite{Loiacono26} studied only a small field (a few pointings of JWST/NIRCam).
The same field, as part of larger data set, was studied by \cite{Kapoor26}.
They reported substantially lower number densities.
Since the results from larger footprint are closer to the general trend found in the literature and our fit, it suggests the observational bias or cosmic variance rather than lack of the evolution.
The slow evolution begins between $z>4$, where number densities reach $\sim 8\times10^{-5}$, and $z\sim2$, dropping to $\sim10^{-6}$.
This is a bit later in cosmic time, comparing to empirical model proposed by \cite{Inayoshi25}, who studied high-$z$ sample reported by \cite{Kocevski2025}.
Our measurements with VIPERS show the abrupt drop happened only at $z\lesssim1$, by two orders of magnitude.
The next step in LRD astrophysics is understanding what drives this evolution.

\begin{table}[]
    \centering
    \caption{Number densities of V-LRDs.}
    \begin{tabular}{lc}
         \hline
         Redshift range & $N_d$ $[10^{-7}/$Mpc$^3]$ \\
         \hline
         $0.50\leq z<1.00$& $2.9^{+3.0}_{-1.8}$ \\
         $1.00\leq z<1.75$& $7.1^{+3.2}_{-2.1}$ \\
         $1.75\leq z<2.60$& $0.2$\phantom{$^{+0.0}$} \\
         \hline
    \end{tabular}
    \label{Tab:numberdensities}
\end{table}

\subsection{Environments of LRDs}
\begin{figure}[]
\includegraphics[width = 0.48\textwidth]{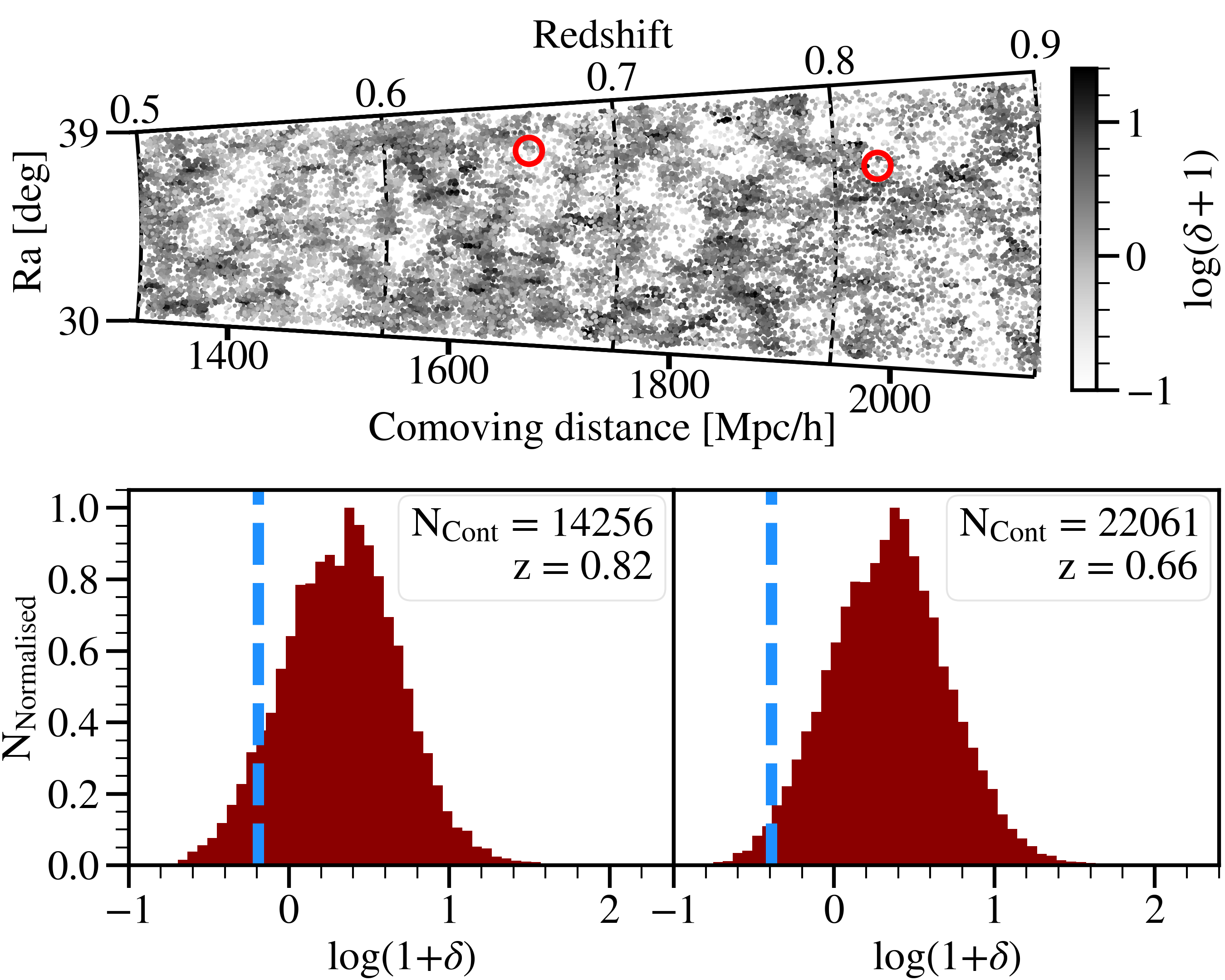}
\caption{Local density of V-LRDs at $z<1$. \textit{Top:} The distribution of local density within VIPERS-CFHTLS W1 strip. The red circles mark the local environment ($\sim10$\,Mpc) of V-LRDs. 
\textit{Bottom:} The red histograms show the local density for control samples, and blue line marks the local density of V-LRD.}
\label{fig:localdens}
\end{figure}

If the obscuring material indeed forms a compact ionized cocoon surrounding the accreting black hole, as recently proposed \citep[][]{Rusakov26, Sneppen26a}, then the origin and long-term supply of this gas become key questions.
To place our V-LRDs in an environmental context, and check if it may help understand its role in the formation of LRD-like objects, we investigate the measured density contrast.
We use only two of V-LRDs at $z<0.9$, as VIPERS local density catalogue is complete up to this redshift (see Section~\ref{sec:datadensity}).
We compare our galaxies with a control sample of VIPERS galaxies matched in redshift ($|z_i - z_\textrm{LRD}|\leq0.1$).
This approach allows us to minimise biases introduced by the dependence of the environment on cosmic epoch.
Since we do not know much about the hosts of LRDs, we do not match the control sample in stellar mass.
The results are presented at Figure~\ref{fig:localdens}.

Both LRD-like objects lie near the edge of a low-density, void-like regions. 
This is qualitatively consistent with higher-$z$ studies \citep[e.g.][]{Carranza-Escudero25}, and a scenario in which the obscuring cocoon is sustained through the accretion of relatively pristine gas from the surrounding intergalactic medium, as low-density environments are less affected by interactions and environmental processing \citep[][]{keres05}.
Thus, such scenario should be available at each redshift.
However, in \cite{Zhang26}, the authors reported a weakening of the relative environmental preference towards lower redshifts.
Since our result concerns two individual systems, larger samples must be established to study population preference.

The current generation of LRD models has been developed primarily to explain the high-redshift population. 
Whether the same physical mechanisms operate at $z\leq1$ remains uncertain. 
The environments of our two analogues therefore provide an additional observational constraint on the conditions under which the LRD phenomenon can occur.
Given that our conclusions are based on only two objects, the present results should be regarded as suggestive rather than conclusive, motivating future studies with larger samples of LRD-like galaxies.

\subsection{X-ray-FIR SED of V-XRD}
In this section, we discuss our attempt to model the SED of V-XRD with Code Investigating GALaxy Emission v25.1 \citep[CIGALE,][]{Boquien19, Yang22}.
The CIGALE uses energy budget to model UV-NIR and NIR-FIR parts simultaneously.
This ensures consistency between stellar populations, dust, and other components.
Finally, the consistency with X-ray emission is provided by the ratio of monochromatic luminosities of the AGN model between X-ray and UV or X-ray and IR (depending on the used module).

We first constrain the photon index ($\Gamma$) from observed data. 
We further use $\Gamma$ as prior in CIGALE to constrain the AGN model and to utilise CIGALE more on UV-FIR  part of SED, focusing on the energy budget.
As this parameter may be affected by other models, especially when relative uncertainties of X-ray fluxes are greater than others \citep[][]{Yang20}, it is a safe solution, which provides a robust $\Gamma$ estimate and also speeds up the modelling.

The $\Gamma$, was estimated by fitting a power-law model to the observed X-ray flux density measurements in logarithmic space, with an assumption that the X-ray spectrum follows:
\begin{equation}
    F_E \propto E^{1-\Gamma},
\end{equation}
where $F_E$ is the flux density per unit of energy and $E$ is the pivotal energy ($E\equiv\sqrt{(E_1\times E_2)}$, where $E_1$ and $E_2$ are the edges of the filter).
The fit was performed using observed EPIC X-ray flux densities and their uncertainties (5 energy bins).  
This approach provides an estimate of $\Gamma=1.38\pm0.16$.
This is a much harder spectrum than the one reported in XRD by \citet[][they measured $\Gamma=1.8$]{Hviding26}, suggesting a different processing of the emission.

\subsubsection{CIGALE modelling of the XRD}
We take advantage of the great UV-FIR coverage of observed (and rest-frame) wavelength by fitting the total SED.
We attempted to model both the host (stars and dust) and the AGN at the same time.
The parameter grid is presented in Appendix~\ref{app:cigaleparam}. 
We use a delayed star formation history (SFH), with possible recent burst; simple stellar populations (SSPs) by \citet[][]{Bruzal03} of solar metallicity and \cite{Chabrier03} initial mass function (IMF); attenuation law by \cite{Charlot00}; dust emission by \cite{Dale14}; and SKIRTOR AGN \citep[][]{Stalevski16}.
On top of that, we test two X-ray modules by \cite{Yang20} and \cite{lopez24} with semi-fixed $\Gamma$ (we allow only 3 values within 1$\sigma$ from our estimation).

\begin{figure}[]
\includegraphics[width = 0.48\textwidth]{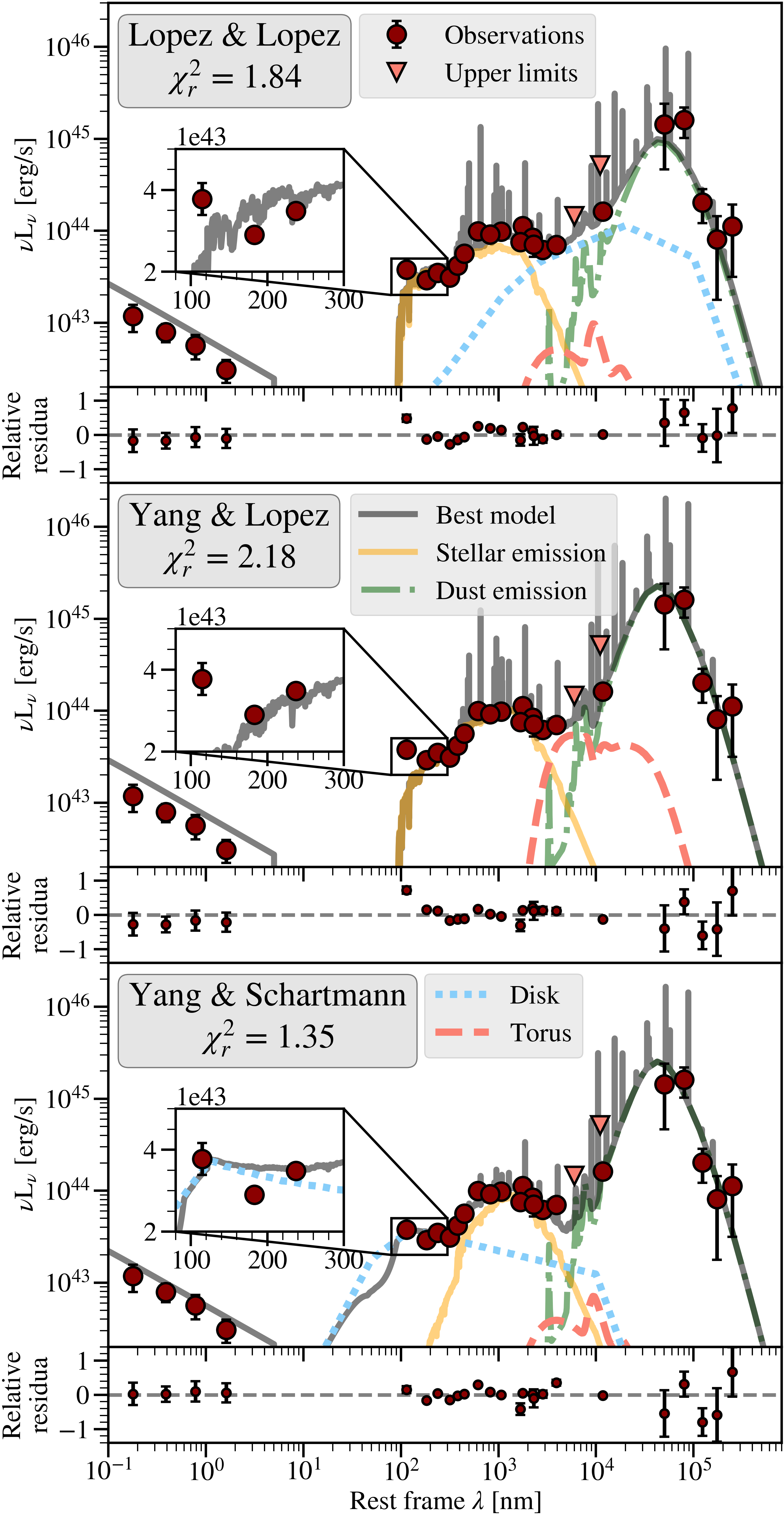}
\caption{Three models of the full SED of V-XRD. From top: LL model, YL model, and YS model. Below each model, we present the relative residua: (observation-model)/observation. The red circles mark the observed fluxes, while the light-red triangles show the upper limits. The solid gray line present the best model. We also present the emission of individual components: stellar attenuated, AGN disk, AGN torus, and dust as solid orange line, dotted blue line, dashed red line, and dot-dashed green line, respectively.}
\label{fig:xrd_sed}
\end{figure}

In total, we conduct three runs: 1) with both X-ray module and disk type introduced in \cite{lopez24}, hereafter LL; 2) with X-ray module by \cite{Yang20} and disk model by \cite{lopez24}, hereafter YL; 3) with X-ray model by \cite{Yang20} and disk model by \cite{Schartmann05}, hereafter YS. 
The first model allows us to compare directly with analysis proposed by \cite{Hviding26}.
The latter two test whether any X-ray/disk combination can reproduce the entire shape of the SED.
The exact parameter spaces are presented in Appendix~\ref{app:cigaleparam}.
The results of the most important parameters are reported in Table~\ref{Tab:cigResult}.
Finally, we show the best fit for each model at Figure~\ref{fig:xrd_sed}.

First of all, each model suggests a quite massive starburst host galaxy, depending on model $\log($M$_\star/$M$_\odot)\sim10.6-11.0$ and SFR$\sim${111-154 M$_\star$/yr}.
We compared the V-XRD with the main sequence (MS) using the MS distance defined as $\Delta$MS$\equiv\log(\textrm{SFR/SFR}_\textrm{MS})$, which is a widely used criterion for selecting star formation outliers \citep[see e.g.][]{Elbaz18, Donevski20, Lisiecki26}.
The models consistently point to a starburst galaxy.
For the MS proposed by \cite{Koprowski24} we report $\Delta$MS$\geq0.5$ dex, 
while for \cite{Popesso23} MS we report $\Delta$MS$\geq0.43$ dex.
A follow-up observation with higher spatial resolution would be required to fully test the starburst scenario.

All three models yield a low AGN fraction, with f${_{\rm AGN}\leq0.04}$. 
However, this should not be interpreted as evidence for a weak AGN. 
In CIGALE, f$_{\rm AGN}$ is defined as the fraction of the total IR luminosity attributed to the AGN dusty torus, rather than the fraction of the bolometric luminosity produced by the accretion-powered component. 
Thus, a powerful AGN can still have a low f$_{\rm AGN}$ if its torus contributes relatively little to the IR emission. 
In our case, the host galaxy is highly obscured and contributes substantially to the infrared emission, while the AGN does not show a strong torus component (see Figure~\ref{fig:xrd_sed}), naturally resulting in the low inferred f$_{\rm AGN}$. 
A low f$_{\rm AGN}$ was also reported for the XRD source by \citet[][f$_{\rm AGN}\sim0.1$]{Hviding26}.

Each model favours a low opening angle of the dusty torus (the angle between equatorial plane and the edge of a torus, $oa$).
This results in only a small portion of the sphere around the central engine covered by the dusty structure. 
With such geometry, the YL model due to large inclination ($i$), is an obscured AGN, which also results in negligible disk emission.
Both, LL and YS, are probing more face-on geometry, looking into unobscured engine, resulting in visible disk contribution. 
However, LL disk component is dominated by IR emission and its UV contribution is not large enough. 
This is consistent with results presented by \cite{Hviding26}, where a similar parameter grid was used. 
However, we find that the YS disk is strong enough to explain observed fluxes.
This is directly reflected with the $\chi^2_r$, which favours the YS model.
This scenario does not require the additional gaseous structure proposed by LRD studies nor the patchy multiphase component proposed to explain XRD by \cite{Hviding26}.
These results suggests that objects selected through similar UV-to-optical SED shapes may nevertheless exhibit substantially different accretion and obscuration properties \citep[see also][]{Chen26}.

The preference for the YS model is primarily driven by the strong rest-frame far-UV emission. 
The \cite{Schartmann05} disk model assumes a radiatively efficient, optically thick accretion disk and therefore produces a strong UV continuum.
The \cite{lopez24} prescription allows for a radiatively inefficient, truncated-disk component with substantially weaker UV emission.
This class of AGN models was constructed to address low-luminosity AGN SEDs \citep{Yuan2004,Nemmen2014}. V-XRD is brighter than LLAGN, as its $\nu L_{\nu}(2\,\textrm{keV}) \approx 7\times 10^{43}$ erg s$^{-1}$ place it just below the distribution of bright X-ray quasars \citep{Chira26}.  
Consequently, the LL combination cannot simultaneously reproduce the observed far-UV and X-ray emission, yielding an extremely X-ray-dominated intrinsic spectrum and a poorer fit compared with YS. 
This suggests that the observed SED is more consistent with a radiatively efficient accretion disk than with the radiatively inefficient accretion-flow scenario. 

\begin{figure}[]
\includegraphics[width = 0.48\textwidth]{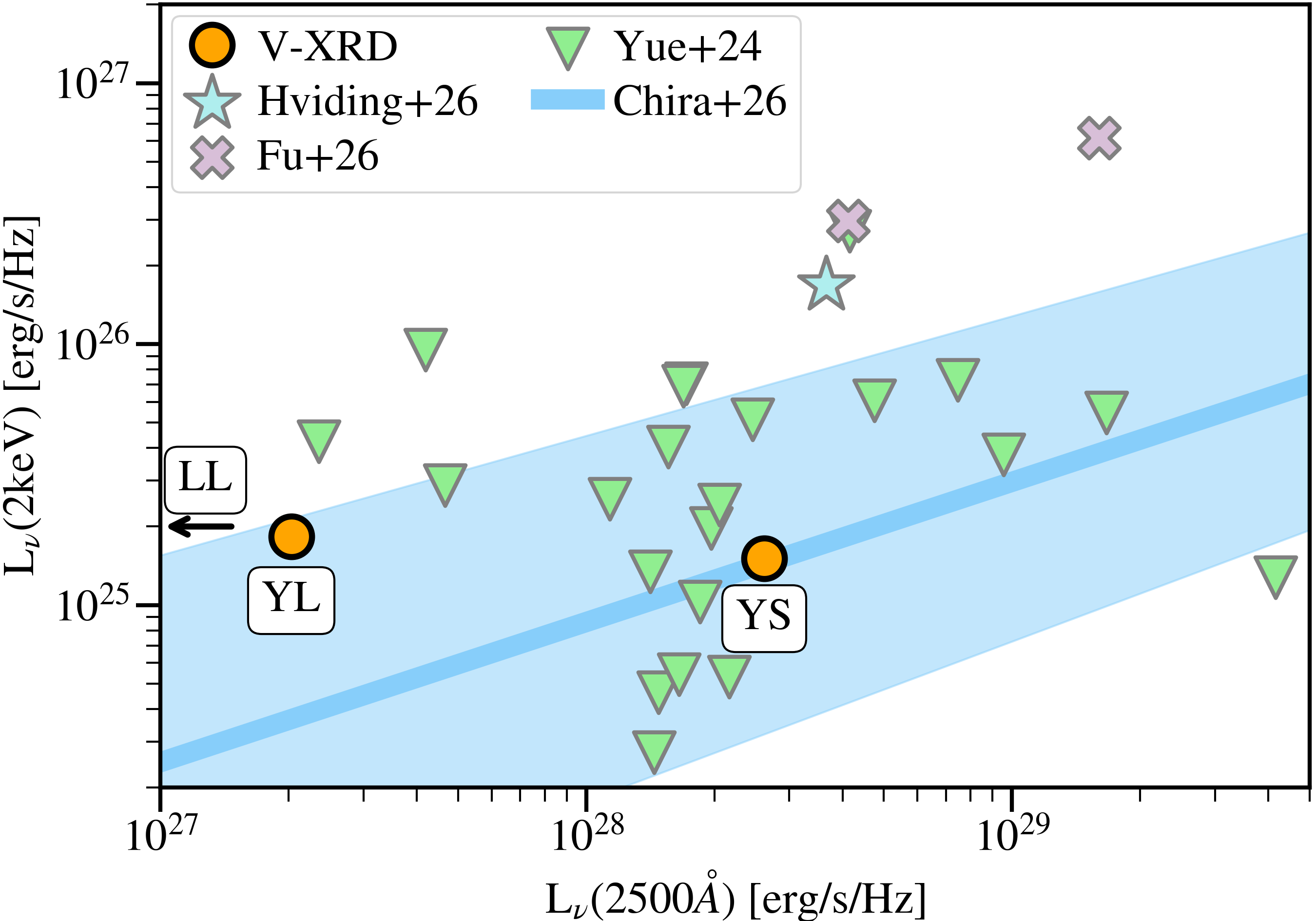}
\caption{$L_{\textrm{2keV}}$ vs $L_{2500\AA}$ for LRD-like objects in literature. We show V-XRD CIGALE models in orange circles. The LL model has ${L_{2500\AA}\sim10^{23}}$ erg s$^{-1}$ Hz$^{-1}$ and did not fit into the image. The crosses present Forge sources from \cite{Fu26}, while the star show the XRD from \cite{Hviding26}. The triangles indicate upper limits for LRD non-detections reported by \cite{Yue24}. The solid line and shaded region show the standard AGN relation derived for $0.5\leq z\leq3$ by \cite{Chira26}.}
\label{fig:L2500L2kev}
\end{figure}

This is further supported by $\alpha_\textrm{ox}$, effectively probing the ratio of X-ray to UV luminosity, in LL model suggests an extremely X-ray loud source.
Although a similar result was presented in XRD and Forges \citep[][]{Fu26, Hviding26}, in our case the loudness is much more extreme.
The X-ray luminosity of V-XRD is $\sim3$ orders of magnitude larger than the intrinsic luminosity at 2500\ \AA. 
The other two of our models suggest more usual AGN \citep[within scatter presented by][]{Chira26}.
We present this comparison at Figure~\ref{fig:L2500L2kev}.
The YS model gives a consistent result, the overall SED shape also agrees with Type I AGN. 
As a consequence, fairly normal broad line region (BLR) would be possible in contrary to LLAGN/LINER case. 
The presence of a broad \ion{Mg}{II} $\lambda$2800 \AA \  allows for a tentative estimate of the virial mass of the SMBH. Using the fitted width of \ion{Mg}{II} and the continuum luminosity, we obtained { $M_{\textrm{SMBH}} {\sim 10^{8} M_{\odot}}$} using the relation of \cite{Wang2009}. This gives an Eddington ratio of the order of 0.01 for the fitted luminosities, but is still fully consistent with the hard SED. 
Thus, the source can be not far above the transiting accretion rate regime towards LLAGN \citep[ie.][]{Liu2020}. 
This kind of transition would result in a rapid decrease in X-ray and disk luminosities of at least the order of magnitude. 
The lower mass and higher accretion rate at the similar luminosity level most likely resulted in narrower emission lines and a systematically softer SED. 
Whether we should expect visible [\ion{Ne}{V}] $\lambda$3427 \AA \ is unclear, as even bright unobscured X-ray AGNs do not always show this line \citep{Reiss2025}.  

\begin{table}[]
    \centering
    \caption{Results of CIGALE fitting for each model.}
    \begin{tabular}{lccc}
    \hline
          & LL & YL & YS  \\
         \hline
        $\chi^2_r$ & 1.84 & 2.18 & 1.35 \\
        $\log($M$_\star$/M$_\odot$) & 10.61 $\pm$ 0.14 & 10.71 $\pm$ 0.08 & 10.97 $\pm$ 0.11 \\
        SFR [M$_\odot$/yr] & 154 $\pm$ 20 & 149 $\pm$ 27 & 111 $\pm$ 27 \\
        f$_\textrm{AGN}$ & 0.01 $\pm$ 0.01 & 0.04 $\pm$ 0.01 & 0.01 $\pm$ 0.01 \\
        $i$ [deg] & 31 $\pm$ 20 & 88 $\pm$ 7 & 34 $\pm$ 16 \\
        $oa$ [deg] & 37 $\pm$ 17 & 20 $\pm$ 3 & 29 $\pm$ 15 \\
        age [Myr] & 3700 $\pm$ 950 & 3400 $\pm$ 700 & 2500 $\pm$ 1000 \\
        $\alpha_\textrm{ox}$ & -0.79$^\textrm{\textdagger}$ & -1.13 $\pm$ 0.09 & -1.29 $\pm$ 0.05 \\
        $\delta$ & 0.32 $\pm$ 0.16 & 0.23 $\pm$ 0.14 & 0.35 $\pm$ 0.17 \\
        $t$ & 6.74 $\pm$ 3.30 & 3.64 $\pm$ 1.88 & 6.46 $\pm$ 3.46 \\
         \hline
    \end{tabular}
    \label{Tab:cigResult}
    \tablefoot{$^\textrm{\textdagger}$ We do not provide error of $\alpha_\textrm{ox}$ as it was estimated directly on the best model.
        \\
    Each column represents one of our models as described in the text. Rows from top: $\chi^2_r$ is reduced $\chi^2$; $\log($M$_\star$/M$_\odot$) is the stellar mass; SFR is star formation rate; f$_\textrm{AGN}$ is AGN fraction; $i$ is inclination; $oa$ is opening angle; age is look-back time of formation age, i.e. the time since the onset of star formation; $\alpha_\textrm{ox}$ is the ratio of monochromatic luminosity at 2~keV and 2500~\AA; $\delta$ is power-law index for the optical slope of the disk (for YL and YS models) or a tuning parameter for the disk type from 0 (advection-dominated accretion flow) to 1 (thin disk); and $t$ is optical depth at 9.7$\mu$m.}
\end{table}

\section{Summary and conclusion}
We presented a unique set of 13 LRDs and one XRD candidate at previously unexplored redshift, $0.5\geq z \geq2.6$, selected from a homogenous and complete sample of VIPERS. 
We selected them using three most important features characteristic for LRDs in literature, namely: (1) broad line component in the spectra; (2) V-shape SED; (3) unresolved morphology.
In addition, we remove sources with a detection of [\ion{Ne}{V}] line, which traces Type~I AGN. 

We further discussed the properties of our LRDs, XRD, and attempt to place them into the current picture. 
Our main results can be summarised as follows:
\begin{itemize}
    \item We find a clear evolution in number density of LRDs through cosmic time: $\log(N_d) = (-5.8\pm0.5)/(1+z) -3.8\pm0.1$. All of the recent discoveries of LRDs at $z<3$ strongly support this decay, but the physical driver is still unclear.
    \item We investigated the local environment of two LRD candidates at $z<1$ with density contrast. We find both sources occupy an underdense environments, consistent with high$-z$ studies.
    \item We find that in the XRD, the radiatively inefficient disk does not produce strong UV emission, even when observed at low inclination. Our observations are more consistent with a radiatively efficient accretion disk. It does not require a peculiar emitting structure.
    \item We find the averaged SED of all our LRD candidates matches well the previously reported samples. We see a blue UV and red optical SED, although the optical slope is not as steep as in many other studies. We believe it is an observational effect.
\end{itemize}

One important caveat of our selection is the spatial resolution. 
To confirm truly compact nature of our candidates, new observations from space-based observatory are required.
Additionally, homogeneous observations of H$\alpha$ would strengthen the results, as most of our sources lack Balmer line observations, which is an important observational tracer of LRDs.
Both will be provided by \textit{Euclid}. 

\begin{acknowledgements}
    We thank the anonymous referee for the careful reading of the manuscript and the useful comments improving the readability and clarifying the message.
    The authors would like to thank Anna de Graaff for a useful discussion.
    This research was funded in whole or in part by the National Science Center, Poland (grants no. UMO-2023/49/N/ST9/00746, 2023/50/E/ST9/00383, UMO-2024/53/B/ST9//00230, and 2023/49/B/ST9/00066, UMO-2023/51/D/ST9/00147, UMO-2022/47/D/ST9/00419).
    We acknowledge the support of the Polish Ministry of Education and Science through the grant PN/01/0034/2022 under `Perły Nauki' programme.
    We acknowledge support from the GEELSBE2 project with reference PID2023-150393NB-I00 funded by MCIU/AEI/10.13039/501100011033 and the FSE+, and the Consolidación Investigadora IGADLE project with reference CNS2024-154572, and also gratefully acknowledge financial support of the Department of Education, Junta de Castilla y Le\'on and FEDER Funds (Reference: CLU-2023-1-05).
    For the purpose of Open Access, the author has applied a CC-BY public copyright license to any Author Accepted Manuscript (AAM) version arising from this submission.
\end{acknowledgements}

\bibliographystyle{aa}
\bibliography{main} 

\begin{appendix}
\section{GALEX photometry}\label{app:galex}
\FloatBarrier
The $\beta_\textrm{UV}$ of sources for which no GALEX fluxes were reported in the main catalogue nor counterparts we found in GR6+7 AIS catalogue (11 galaxies) could not be estimated.
Thus, we download the background-subtracted images from GALEX via {\tt astroquery.mast} python package \citep[]{astroquery19}.
We select the observations with the longest exposure time for both FUV and NUV filters. 
Some of the observations provided us with multiple exposures. 
We produced cutouts of size 1$\times$1 arcmin around the sources.
If several exposures were available, we reprojected multiple observations to the same grid and stack the images to make one cutout.

We measure the flux with two methods.
First we put a circular aperture of 5 pixels in radius (7.5 arcsec), which is larger than the PSF, and measure the flux.
We also measure the apertures of 3 pixels in radius (4.5 arcsec), which is close to the PSF size.
The error is estimated as the standard deviation of the flux measured in 10 apertures with the same size at positions devoid of emission.
For the second method, we use GALFIT \citep[][]{Peng10} and fit a Gaussian-like PSF of FWHM set to 3 pixels (close to actual PSF size of GALEX observations). 

\begin{figure}[h]
\includegraphics[width = 0.48\textwidth]{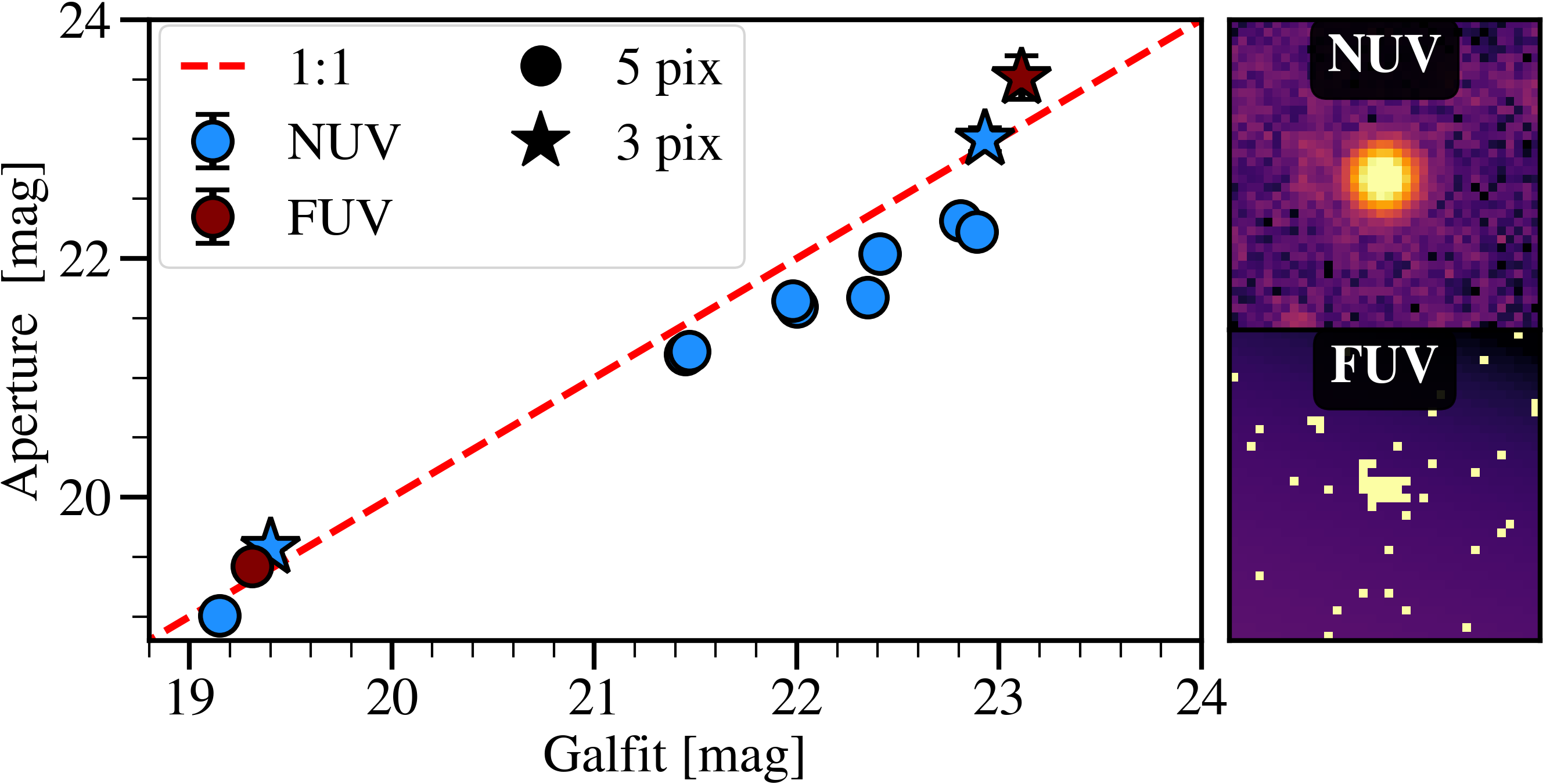}
\caption{\textit{Left: }Comparison of GALFIT and aperture photometry. The blue and red points show the measured fluxes of NUV and FUV, respectively. The red line shows the one-to-one relation.
\textit{Right: } Cutouts of the brightest source that did not have FUV and NUV measurements in the catalogue.}
\label{fig:photometryGalex}
\end{figure}

We compare the results in Figure~\ref{fig:photometryGalex}.
As the fluxes can be up to 0.7 magnitude apart, we use the central value between the aperture photometry and GALFIT fit for each galaxy, and the standard deviation of the empty apertures as the error.
The summary of the additional detections is presented in Table~\ref{Tab:galex}.

\begin{table}[]
    \centering
    \caption{The summary of additional detections with GALEX. Measurements are in magnitude.}
    \begin{tabular}{ccccc}
    \hline
         VIPERS ID & NUV & NUV error & FUV & FUV error  \\
         \hline
         103190305 & 19.079 & 0.004 & 19.366 & 0.022  \\
         107105125 & 22.563 & 0.071 & 23.312 & 0.182  \\
         111199146 & 22.558 & 0.025 & -- & --  \\
         120111832 & 21.799 & 0.053 & -- & --  \\
         401161203 & 22.963 & 0.099 & -- & --  \\
         402180898 & 19.490 & 0.007 & -- & --  \\
         402225508 & 21.324 & 0.034 & -- & --  \\
         403115375 & 22.225 & 0.065 & -- & --  \\
         406060307 & 22.013 & 0.061 & -- & --  \\
         408023687 & 21.812 & 0.041 & -- & --  \\
         408065924 & 21.345 & 0.028 & -- & --  \\
         \hline
    \end{tabular}
    \tablefoot{We note, none of these sources are present in the final sample.}
    \label{Tab:galex}
\end{table}

\FloatBarrier

\FloatBarrier

\section{Morphology fitting}\label{app:morpho}
To find if the V-shaped sources are unresolved, we perform a 2D fit of the light distribution using GALFIT \citep{peng02, Peng10}.
Since we know that these sources host AGNs, we fit two separate models: 1) point source; 2) two-component model, consisting of S\'ersic profile and a point source.
It is worth to note that the model is fitted only after being convolved with the PSF.
Thus, any source of size smaller than the PSF will be effectively treated as point-like.

Before running GALFIT, the images are masked in order to fit the source of interest.
To do so, we first run Source Extractor \citep{Bertin96} to find all sources in the cutout.
Then, using the parameters A\_IMAGE, B\_IMAGE, KRON\_RADIUS and THETA\_IMAGE, we mask all sources but the studied one in the centre.
For each detected source, the mask is an ellipse characterized by SExtractor output parameters.
To ensure that all sources, including the smallest ones, are totally masked, the KRON\_RADIUS of each ellipse were multiplied by 3 and 2 pixels were added (see examples of masks at Figure~\ref{fig:galfitFits}). 
Since we study morphology in two bands, \textit{i} and \textit{y}, we combine both masks to build the master mask for each source.

It is known that GALFIT usually underestimates the errors of returned parameters and may be biased towards local minimum of the parameter space, especially when used to fit models with many parameters \citep{Haussler07}, such as a two-component model with 10 parameters.
Thus, since we want to estimate real uncertainties and ensure the best possible solution, we performed 100 independent GALFIT optimisations per band for each galaxy using identical science images, noise maps, masks, point spread functions, and model configurations.
The only quantity varied between runs was the initial set of model parameters, allowing the optimiser to explore different regions of the parameter space and converge to potentially distinct local minima \citep[similar to e.g.][]{Lange16}.
The initial guesses and the variation ranges are presented in Table~\ref{Tab:galfit}.
The best-fitting solution was identified as the model with the minimum $\chi2$.

\begin{table}[]
    \centering
    \caption{The initial guesses for 2-component model GALFIT fitting.}
    \begin{tabular}{ccc}
    \hline
         GALFIT parameter & Initial guess & Unit \\
         \hline
         \multicolumn{3}{c}{S\'ersic profile}\\
         \hline
            magnitude & $0.25F_\nu\pm0.1$ & mag\\
            semimajor axis $a$ & 3 $\pm2$ & pix\\
            S\'ersic index $n$ & 3 $\pm2$ & --\\
            axis ratio $q$ & 0.7 $\pm0.2$ &--\\
            position angle  & random & deg\\
            position $x_1,y_1$ & centre of image $\pm2$ & pix\\
         \hline
         \multicolumn{3}{c}{Point source}\\
         \hline
            magnitude & $0.75F_\nu\pm0.1$ & mag\\
            position $x_2,y_2$& $x_1,y_1$ & pix\\
         \hline
         \multicolumn{3}{c}{Constrains}\\
         \hline
            S\'ersic index & 0.2-6 & --\\
            axis ratio & 0.1-1 & --\\
            position of point source & $x_1\pm 0.5, y_1\pm 0.5$  &pix\\
         \hline
    \end{tabular}
    \tablefoot{The $F_\nu$ is the observed flux density in the catalogue, later transformed to AB magnitude to match GALFIT input units.}
    \label{Tab:galfit}
\end{table}

\begin{figure}[h]
\includegraphics[width = 0.48\textwidth]{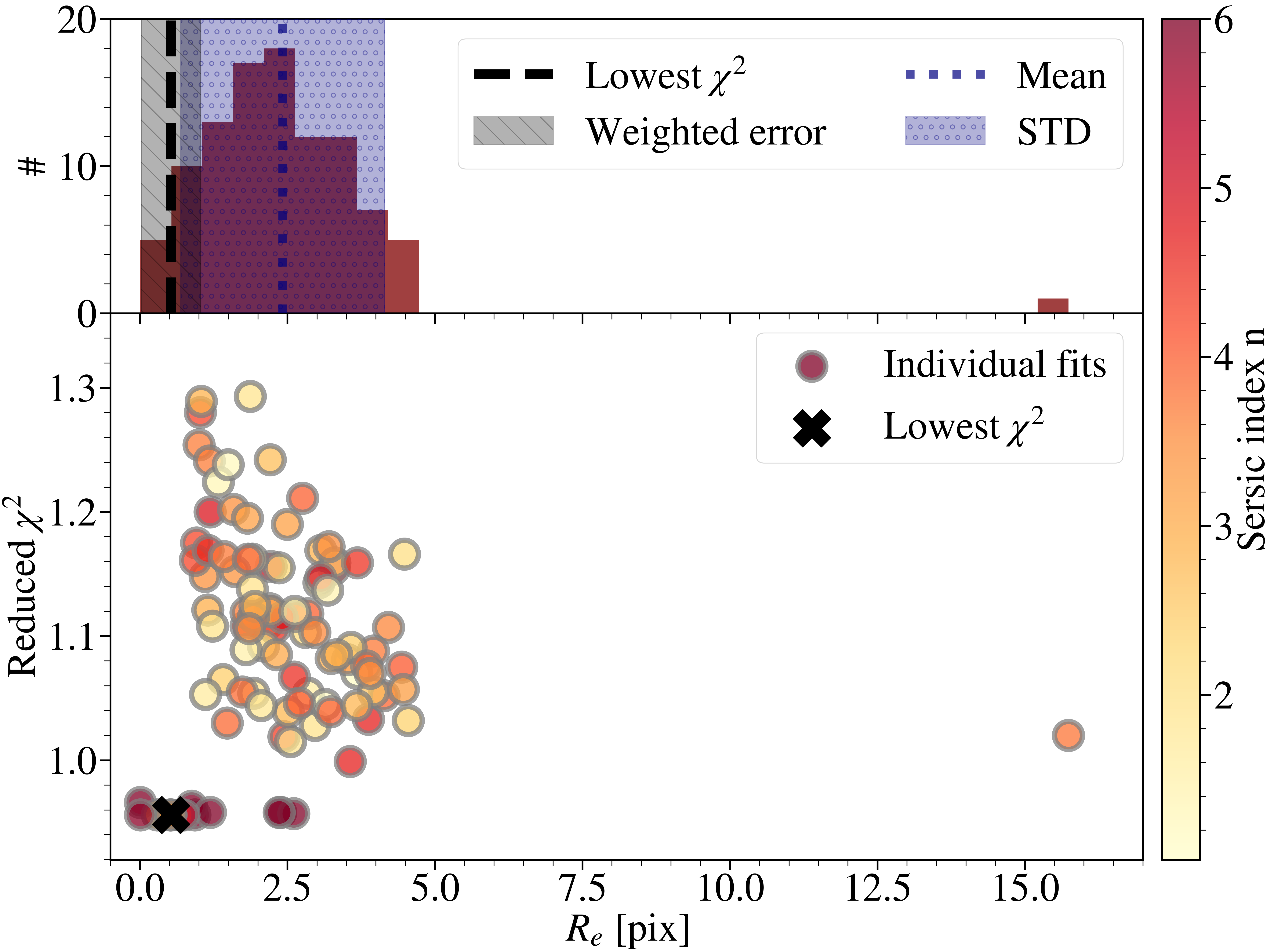}
\caption{Exemplary results of 100 fits with GALFIT and comparison of the error propagation. \textit{Bottom:} Reduced $\chi^2$ vs $R_e$. Individual runs are presented with circles. Colour corresponds to the S\'ersic index. We denote the best fit with black cross. \textit{Top:} The histogram of $R_e$ results. We mark the best fit with dashed black line and the $1\sigma$ weighted error with the shaded grey region. For comparison, we show the mean value with dotted dark blue line and the the standard deviation with dark blue shaded region.}
\label{fig:morpho_error}
\end{figure}

We further estimate the uncertainty of each parameter.
Rather than adopting the distribution of all converged solutions directly, which would over-represent poor local minima, we assigned each solution a likelihood weight based on its relative goodness of fit.
\begin{equation}
    w_i = \exp\left(-\frac{\Delta\chi_i^2}{2}\right),
\end{equation}
where $\Delta\chi_i^2 = \chi_i^2 - \chi_{\min}^2$. 
The final uncertainty of each fitted parameter was computed as the weighted standard deviation of the ensemble of converged solutions using these likelihood weights, thereby emphasising statistically equivalent solutions while naturally suppressing poor local minima.
The exemplary comparison of such a weighted error with the standard deviation is presented in Figure~\ref{fig:morpho_error}.

\FloatBarrier
\section{Spectral fitting}\label{sec:spectralfitting}
Emission-line and continuum fitting were performed using \texttt{PyQSOFit\footnote{\url{github.com/legolason/PyQSOFit}}} \citep{Guo18, Shen19}, adopting a simultaneous decomposition of the continuum and emission-line components. 
For each spectrum, the fitting wavelength range was restricted to the observed rest-frame range, and continuum windows were selected according to spectral coverage, signal-to-noise (S/N) ratio, and the absence of strong emission features. 
The continuum was modelled with a power-law component. 
Emission lines were fitted as Gaussian components. 
Broad and narrow components were constrained separately using priors on velocity width and centroid offsets consistent with the instrumental resolution of VIPERS ($R\simeq220$). 
Best-fitting parameters were obtained through maximum-likelihood optimisation, from which integrated line fluxes, equivalent widths, full width at half maximum (FWHM), and signal-to-noise ratios were extracted for each emission-line complex. 
The quality of each fit was assessed using the reduced $\chi^2$ computed directly from the fit residuals.

\subsection{Spectrum staking algorithm}\label{app:stack}
\begin{figure}[h]
\includegraphics[width = 0.48\textwidth]{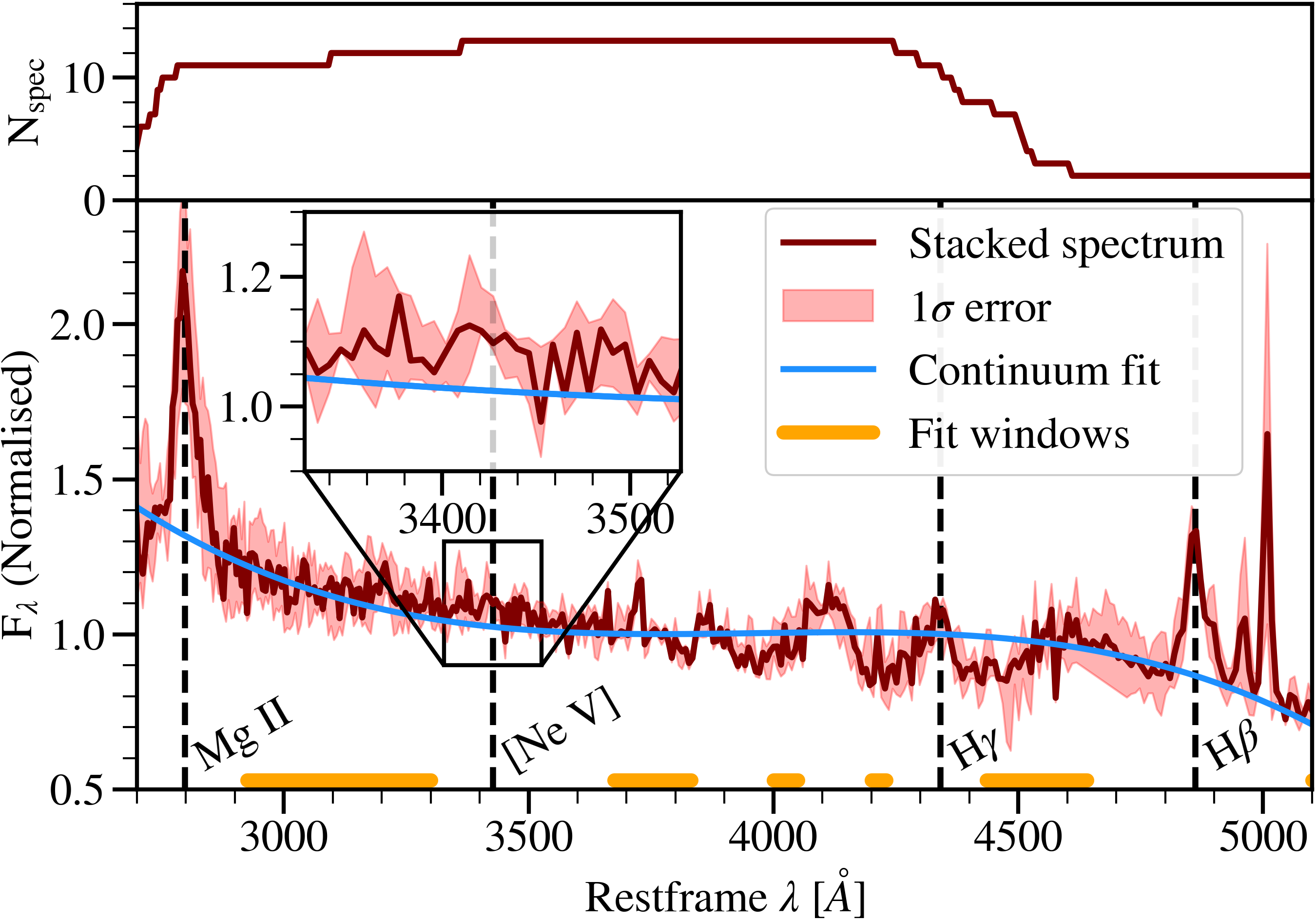}
\caption{\textit{Top:} Number of sources with observation at given rest frame wavelength. \textit{Bottom:} Stacked spectra of final sample (14 objects) in dark red. The blue line shows the continuum fit. We mark the continuum fitting windows in orange at the bottom of the plot. The dashed black lines mark the lines. The inset axis is a zoom-in into [\ion{Ne}{V}] region. }
\label{fig:stackedSpec}
\end{figure}

We stacked the spectra of the 13 objects (we do not stack object with $z>1.75$) within final sample (V-shape SED, compact morphology, and no individual detection of [\ion{Ne}{V}]). 
First, we mask pixels considered not trustable according to VIPERS documentation\footnote{\url{eso.org/rm/api/v1/public/releaseDescriptions/98}}.
We remove all channels with mask 1 and 2 (edited channels).
Additionally, we remove channels with mask 3 and with SNR<5.
We further reprojected the spectra into the same, oversampled grid at rest frame, log-binned wavelength. 
We normalise each spectrum for the continuum level, defined as the median flux value at wavelength range $3675 >\lambda> 3832$ (one of the continuum fitting windows defined in PyQSOFit).
Finally, we stack the spectra by calculating the median in each channel
and find the uncertainties by bootstrapping the sample 2000 times. 
The continuum was fitted on the stacked spectrum as a polynomial using \texttt{PyQSOFit}.
The result is presented in Figure~\ref{fig:stackedSpec}.

\FloatBarrier

\section{Final sample presentation}\label{App:finalPresent}
\begin{figure*}[]
\includegraphics[width = 0.98\textwidth]{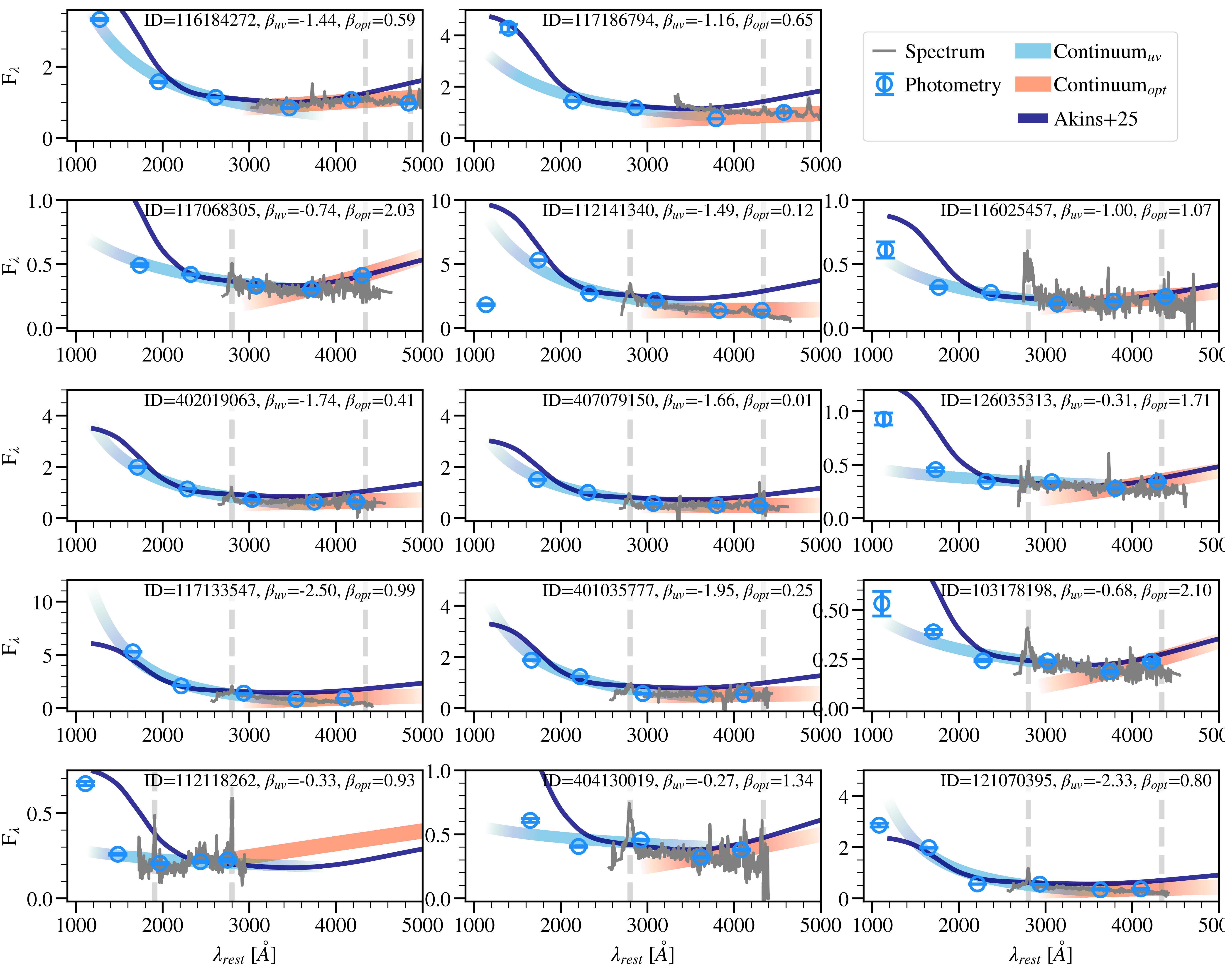}
\caption{SEDs of objects in final sample. The solid gray line shows the scaled spectrum, the blue points represent the broadband observations with errors. 
We plot the optical and UV slopes with light-red and light-blue, respectively. 
The dark blue solid line shows the normalised stacked SED of high-$z$ LRD sample studied by \cite{Akins25}.
We mark the lines considered to be broad with light-gray dashed line.}
\label{fig:sed1}
\end{figure*}

\begin{figure*}[h]
\includegraphics[width = 0.98\textwidth]{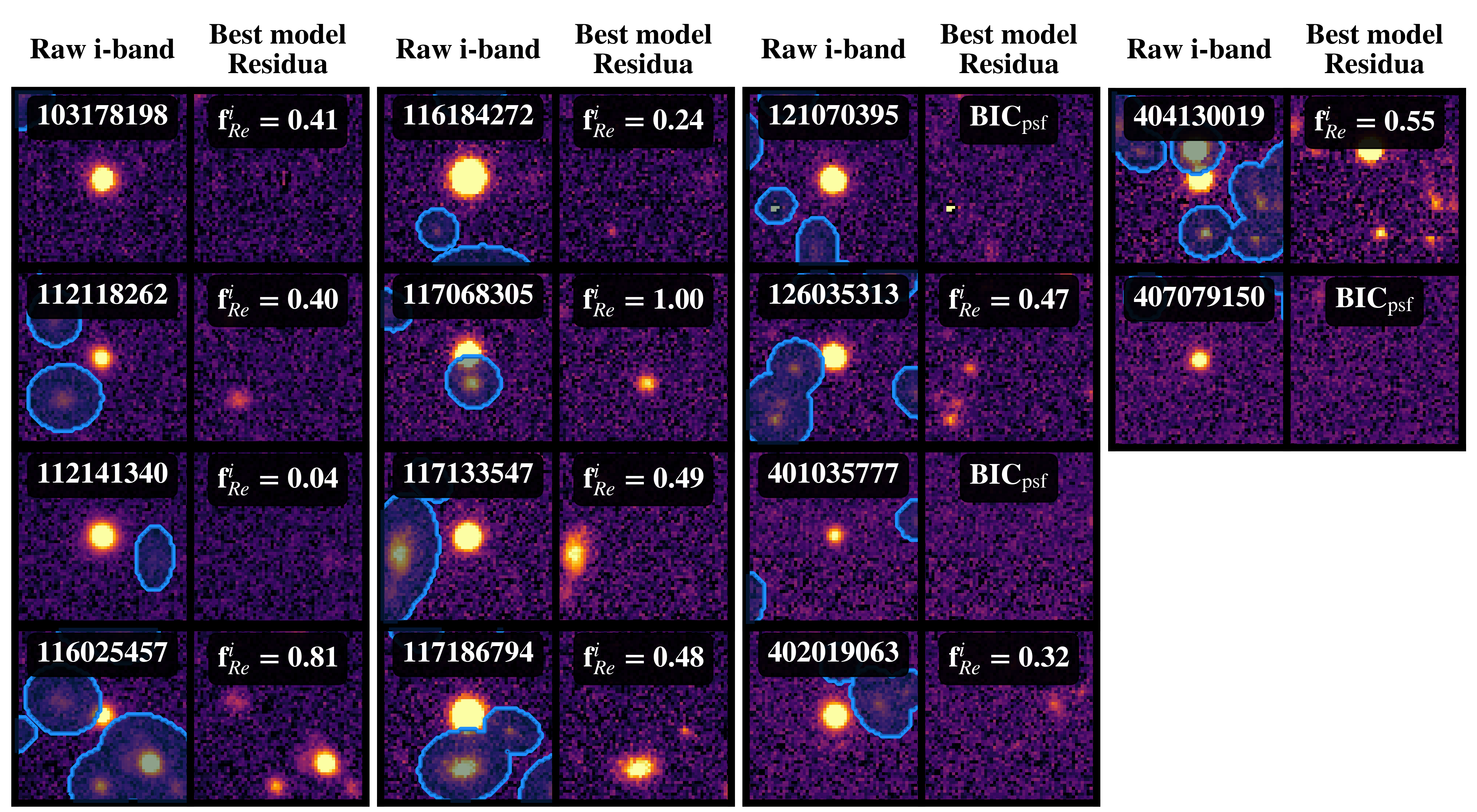}
\caption{Morphology fits for each galaxy in our final sample. We show the raw $i$-band with mask as blue shaded region in the left column, and residua from best model in the right column. The VIPERS ID is given for each source in the left column. Additionally, the ratio f$_{Re}\equiv(R_e+\Delta R_e)/R_e^\textrm{psf}$ for $i$-band (or an indicator for BIC criterion) is given in the right column.}
\label{fig:morpho_all}
\end{figure*}

We report the properties of galaxies in final sample in Table~\ref{Tab:finaltable}.
The SED of each source is presented in Figure~\ref{fig:sed1}.
Additionally, we present the $i$-band cutouts for each source, together with the best morphology model residua in Figure~\ref{fig:morpho_all}.

\begin{table*}[]
    \centering
    \caption{Summary of measurements for our final sample.}
    \begin{tabular}{c c c c c c c c c c c c}
    \hline
    VIPERS ID & Ra [deg] & Dec [deg]& $z_\textrm{spec}$ & $\beta_\textrm{uv}$& $\beta_\textrm{opt}$ &$R_e^i$  & $R_{e,i}^\textrm{psf}$  & $n_i$ &$R_e^y$ & $R_{e,i}^\textrm{psf}$ & $n_y$ \\
    \hline
117186794 & \phantom{0}37.3081 & -4.7052 & 0.66 & -1.16 & 0.65 & 0.78$\pm$0.07 & 1.8 & 2.8$\pm$0.2 & 1.77$\pm$0.12 & 1.9 & 0.8$\pm$0.2 \\
116184272 & \phantom{0}36.3647 & -4.7095 & 0.82 & -1.44 & 0.59 & 0.30$\pm$0.12 & 1.8 & 5.1$\pm$0.7 & 0.56$\pm$0.17 & 1.8 & 5.7$\pm$1.0 \\
\underline{116025457} & \phantom{0}36.0544 & -5.4933 & 1.01 & -1.00 & 1.07 & 1.53$\pm$0.11 & 2.0 & 3.1$\pm$0.3 & 1.79$\pm$0.12 & 2.4 & 0.2$\pm$0.2 \\
112141340 & \phantom{0}32.6382 & -4.9637 & 1.04 & -1.49 & 0.12 & 0.01$\pm$0.07 & 2.0 & 5.9$\pm$1.1 & 0.51$\pm$0.28 & 2.4 & 0.2$\pm$0.5 \\
117068305 & \phantom{0}37.2892 & -5.2883 & 1.05 & -0.74 & 2.03 & 1.85$\pm$0.03 & 1.9 & 0.8$\pm$0.0 & 1.53$\pm$0.11 & 2.0 & 2.6$\pm$0.8 \\
126035313 & \phantom{0}37.4596 & -4.5010 & 1.05 & -0.31 & 1.71 & 0.86$\pm$0.00 & 1.8 & 2.0$\pm$0.0 & 1.06$\pm$0.01 & 1.9 & 1.2$\pm$0.1 \\
407079150 & 330.4602 & \phantom{-}2.0825 & 1.05 & -1.66 & 0.01 & -- & 1.7 & -- & -- & 1.7 & --  \\
402019063 & 331.2979 & \phantom{-}0.8935 & 1.08 & -1.74 & 0.41 & 0.43$\pm$0.13 & 1.7 & 6.0$\pm$0.2 & 1.30$\pm$0.28 & 1.6 & 0.2$\pm$2.1 \\
103178198 & \phantom{0}32.5113 & -5.7929 & 1.09 & -0.68 & 2.10 & 0.84$\pm$0.02 & 2.1 & 4.3$\pm$0.1 & 1.43$\pm$0.30 & 2.4 & 3.0$\pm$0.6 \\
401035777 & 330.4220 & \phantom{-}0.9720 & 1.14 & -1.95 & 0.25 & -- & 1.7 & -- & -- & 1.5 & --  \\
117133547 & \phantom{0}37.2897 & -4.9845 & 1.15 & -2.50 & 0.99 & 0.68$\pm$0.22 & 1.8 & 0.2$\pm$0.4 & 0.23$\pm$0.37 & 2.0 & 1.8$\pm$1.6 \\
121070395 & \phantom{0}32.1357 & -4.3678 & 1.15 & -2.33 & 0.80 & -- & 2.2 & -- & -- & 2.4 & --  \\
404130019 & 332.8696 & \phantom{-}1.4175 & 1.16 & -0.27 & 1.34 & 0.80$\pm$0.03 & 1.5 & 4.5$\pm$0.4 & 1.91$\pm$0.07 & 2.1 & 0.5$\pm$0.1 \\
112118262 & \phantom{0}32.3172 & -5.0712 & 2.20 & -0.33 & 0.93 & 0.76$\pm$0.06 & 2.0 & 1.2$\pm$0.1 & 1.66$\pm$0.43 & 2.5 & 0.2$\pm$0.9 \\
    \hline
    \end{tabular}
    \tablefoot{Each column represents one of the important parameters used in the analysis as described in the text. We report optical slope as $\beta_\textrm{opt}$. The lower index next to effective radius ($R_e$) or S\'ersic index ($n$) refers to given HSC band. The $R_e$ is given in pixels.\\
    For galaxies that BIC of point like model is lower than 2-component model, we do not report S\'ersic fit.\\
    We underline the V-XRD source.}
    \label{Tab:finaltable}
\end{table*}

\FloatBarrier

\section{Completeness correction}\label{app:completeness}
To calculate the corrected number of sources in each redshift bin, we account for the probability that a galaxy is observed spectroscopically. 
For the standard VIPERS target sample, this probability is defined as:
\begin{equation}
    P_{\rm obs} = {\rm SSR}\times{\rm TSR}.
\end{equation}
For details see \cite{garilli14}, \cite{guzzo14}, and \cite{scodeggio18}.
However, five sources in our V-LRD sample do not have assigned TSR or SSR values. 
All five were initially classified as possible stellar contaminants and were therefore excluded from the PDR1 target sample. 
Prior to PDR2, objects classified as stellar contaminants were additionally tested against the AGN selection criteria; those satisfying the AGN criterion were subsequently added to the target sample. 
Because these sources were selected through a different selection channel, they were not assigned the standard TSR and SSR values \cite{scodeggio18}. 
We therefore estimate their observation probability empirically by comparing the number of stellar-classified AGN candidates with a VIPERS spectrum, $N_{\rm obs}$, to the total number of such sources in the CFHTLS parent catalogue of \citet[][$N_{\rm par}$]{scodeggio18}. 
This gives:
\begin{equation}
    P_{\rm obs}^{\rm AGN} \equiv
    \frac{N_{\rm obs}}{N_{\rm par}} = 0.32,
\end{equation}
which we us as substitute for the mentioned five sources.

To assess the potential incompleteness caused by the initial stellar classification, we additionally estimate the number of broad-line AGN that may have been missed by the AGN selection. 
We selected sources classified as stellar but not flagged as AGN candidates in the VIPERS parent catalogue; although these sources were not targeted by the survey, a small fraction nevertheless received spectroscopic observations. 
Among these observed sources, we measured the fraction subsequently identified as broad-line AGN, obtaining $f_\star^{\rm AGN}=0.015$. 
Assuming that this fraction also applies to the unobserved stellar contaminants, we estimate the potentially missed broad-line AGN population, $N_{\rm miss}=1593$, corresponding to an estimated missed-to-observed BLAGN ratio of $f_{\rm miss}^{\rm AGN} = 1.28.$.

Finally, we calculate the effective, completeness-corrected number of sources as:
\begin{equation}\label{eq:weightedN}
    N_{\rm w} = \sum_{i=1}^{N} \frac{1}{P_{{\rm obs},i}}.
\end{equation}
For the fiducial estimate, the five sources without assigned TSR and SSR are weighted using $P_{\rm obs}^{\rm AGN}$. 
For the upper-limit estimate, we additionally account for the potentially missed broad-line AGN population associated with stellar contaminants.
We apply the correction factor, $C_{\rm miss} = 1+f_{\rm miss}^{\rm AGN}$, only to the contribution of these five sources, so that their effective observation probability is
\begin{equation}
    P_{\rm obs}^{\rm upper} =
    \frac{P_{\rm obs}^{\rm AGN}}{C_{\rm miss}}.
\end{equation}
For the lower-limit estimate, we conservatively assume that the five sources require no completeness correction and therefore set $P_{\rm obs}=1$.
We additionally propagate Poisson fluctuations in the source counts to obtain the final lower and upper limit estimates.
\FloatBarrier

\section{CIGALE parameters}\label{app:cigaleparam}
We present an input parameter grid for each model in Table~\ref{TAB:CIG}
\begin{table}[ht] 
\centering
\label{TAB:CIG}
\caption{CIGALE parameter space.}
    \begin{tabular}{l r}
    Parameter & grid\\
    \hline
    \multicolumn{2}{c}{Delayed SFH}\\
    \hline
    $\tau_\textrm{main}$ [Myr]& 500, 2000, 7000  \\
    age$_\textrm{main}$ [Myr] & 250, 1688, 3125, 4563, 6000 \\
    $\tau_\textrm{burst}$ [Myr] & 150 \\
    age$_\textrm{burst}$ [Myr] & 50 \\
    $f_\textrm{burst}$ & 0, 0.01, 0.1, 0.2 \\
    \hline
    \multicolumn{2}{c}{SSP \cite{Bruzal03}}\\
    \hline
    IMF & 1 \citep[][]{Chabrier03}\\
    \hline
    \multicolumn{2}{c}{Attenuation \cite{Charlot00}}\\
    \hline
    $A_\textrm{V}\textrm{(ISM)}$ & 0.1, 1.0, 1.5, 1.8, 2, 2.5, 3, 3.5, 4.5\\
    \hline
    \multicolumn{2}{c}{Dust emission \cite{Dale14}}\\
    \hline
    $\alpha$ & 0.125, 0.25, 0.5, 1 \\
    \hline
    \multicolumn{2}{c}{AGN \cite{Stalevski16}}\\
    \hline
    $\tau_{9.8\mu m}$ & 3, 7, 11\\
    $pl$ & 1\\
    $q$ & 1\\
    $oa$ & 20, 40, 60, 80\\
    $R$ & 10, 20, 30 \\
    $i$ & 10, 30, 50, 70, 90\\
    $\delta$ & 0, 0.25, 0.5\\
    fracAGN & 0.001, 0.003, 0.01, 0.04, 0.1, 0.18\\
    EBV & 0\\
    disk\_type &2 (LL), 2 (YL), 1(YS)\\
    \hline
    \multicolumn{2}{c}{X-ray (for LL model) \cite{lopez24}}\\
    \hline
    $\Gamma$ & 1.22, 1.38, 1.54\\
    $\alpha_\textrm{irx}$ & 0, 0.15, 0.3, 0.45, 0.6\\
    \hline
    \multicolumn{2}{c}{X-ray (for YL and YS models) \cite{Yang20}}\\
    \hline
    $\Gamma$ & 1.22, 1.38, 1.54\\
    $\alpha_\textrm{ox}$ & -0.9, -1.1, -1.3, -1.5, -1.8, -2.2\\
    \hline
\end{tabular}
\tablefoot{If a parameter is not explicitly written, the initial values were kept. The disk\_type parameter had only one value for each model.}
\end{table}

\end{appendix}
\end{document}